\documentclass[aps,prc,twocolumn,groupedaddress]{revtex4-2}

\usepackage{graphicx}
\usepackage{slashed}
\usepackage{ulem}
\usepackage{color}
\newcommand{\nopieft}{\mbox{$\slashed{\pi}$EFT~}}

\begin{document}

\title{Nature of the $\Lambda nn$ $(J^\pi=1/2^+, I=1)$ and ${\rm ^3_\Lambda H^*} (J^\pi=3/2^+, I=0)$ states}

\author{M. Sch\"{a}fer}
\email[]{m.schafer@ujf.cas.cz}
\affiliation{Czech Technical University in Prague, Faculty of Nuclear Sciences and Physical Engineering, B\v{r}ehov\'{a} 7, 11519 Prague 1, Czech Republic}
\affiliation{Nuclear Physics Institute of the Czech Academy of Sciences, 25069 \v{R}e\v{z}, Czech Republic}

\author{B. Bazak}
\email[]{betzalel.bazak@mail.huji.ac.il}
\affiliation{Racah Institute of Physics,The Hebrew University, Jerusalem 91904, Israel}

\author{N. Barnea}
\email[]{nir@phys.huji.ac.il}
\affiliation{Racah Institute of Physics,The Hebrew University, Jerusalem 91904, Israel}

\author{J. Mare\v{s}}
\email[]{mares@ujf.cas.cz}
\affiliation{Nuclear Physics Institute of the Czech Academy of Sciences, 25069 \v{R}e\v{z}, Czech Republic}

\date{\today}

\begin{abstract}
The nature of the $\Lambda nn$ and ${\rm ^3_\Lambda H^*} (J^\pi=3/2^+,~I=0)$ states is investigated within a pionless effective field theory at leading order (LO \nopieft), constrained by the low energy $\Lambda N$ scattering data and hypernuclear 3- and 4-body data. 
Bound state solutions are obtained using the stochastic variational method, the continuum region is studied by employing two independent methods - the inverse analytic continuation in the coupling constant method and the complex scaling method. Our calculations yield both the $\Lambda nn$ and ${\rm ^3_\Lambda H^*}$ states unbound. We conclude that the excited state ${\rm ^3_\Lambda H^*}$ is a virtual state and the  $\Lambda nn$ pole located close to the three-body threshold in a complex energy plane could convert to a true resonance with Re$(E)>0$ for some considered $\Lambda N$ interactions. Finally, the stability of resonance solutions is discussed and limits of the accuracy of performed calculations are assessed. 
\end{abstract}
\maketitle

\section{Introduction}
The $s$-shell $\Lambda$ hypernuclei play an important role in the study of 
baryon-baryon interactions in the strangeness sector.
In view of scarce hyperon-nucleon scattering data they provide a unique test ground for the underlying interaction models thanks to  reliable few-body techniques. In particular, experimental values of the $\Lambda$ separation energies in $A=3,\, 4$ $\Lambda$ hypernuclei including their known spin and parity assignments, as well as the $^4_{\Lambda}$H$^*$ and $^4_{\Lambda}$He$^*$ excitation energies represent quite stringent constraints (see \cite{GHM16} and references therein). 

The hypertriton $^3_\Lambda$H ($J^\pi=1/2^+,\, I=0$) is the lightest known hypernucleus, with the $\Lambda$ separation energy $B_{\Lambda} = 0.13 \pm 0.05$~MeV \cite{davis05} used as few-body constraint in the present study. This value, considered well established for decades, has been challenged recently by the STAR collaboration which has reported a value $B_{\Lambda} = 0.41 \pm 0.12$~MeV \cite{star20}. Possible implications of this larger value for hypernuclear calculations have been studied in Refs.~\cite{le20, HH20}. In view of the small value of $B_{\Lambda}$ in the hypertriton ground state, it is likely that the excited state $^3_\Lambda$H$^*$ ($J^\pi=3/2^+,\, I=0$) is located just above the $\Lambda+d$ threshold, however, its physical nature is not known yet. Moreover, since the isospin-triplet $NN$ state is unbound, it is highly unlikely that there exists a bound state in the $I=1$  $\Lambda nn$ system. A thorough study of the $A=3$ hypernuclear systems  with different spin and isospin, addressing the question whether they are bound or continuum states, provides invaluable information about the spin and  isospin dependence of the $\Lambda N$ interaction, as well as dynamical effects in these few-body systems caused by a $\Lambda$ hyperon. Moreover, the issue of the $\Lambda nn$ and also $\Lambda \Lambda nn$ states as possible candidates for widely discussed bound neutral nuclear systems has attracted  increased attention recently in connection with the experimental evidence for the bound $\Lambda nn$ state reported by the HypHI collaboration \cite{HypHI13}. 

The first variational calculation demonstrating that the $\Lambda nn$ system is unbound was performed by Dalitz and Downs more than 50 years ago \cite{DD59}. Later, this conclusion was further supported by Garcilazo using Faddeev approach with separable potentials \cite{garcilazo87}. Following, more detailed, studies of both $\Lambda nn$ and $^3_\Lambda$H$^*$ systems within Faddeev approach using either Nijmegen $YN$ potential \cite{MKGS95} or chiral constituent quark model of $YN$ interactions \cite{GFV07a, GFV07b} confirmed that both systems are indeed unbound. In addition, these calculations revealed that with increasing $YN$ attraction the binding of $^3_\Lambda$H$^*$ comes first. The investigation of the $\Lambda d$ scattering length in $J^\pi=3/2^+$ channel indicated existence of a pole in the vicinity of the $\Lambda+d$ threshold. Continuum calculations of the unbound $\Lambda nn$ system were performed by Belyaev et al. using a phenomenological $\Lambda N$ potential \cite{BRS08}. This neutral hypernuclear system was found to form a very wide, near-threshold resonance.  

In view of the above theoretical calculations, the claimed evidence of the $\Lambda nn$ reported by HypHI Collaboration was quite surprising and it stimulated renewed interest in the nature of the 3-body hypernuclear states. The HypHI conclusions were seriously challenged by succeeding calculations \cite{GG14,HOGR14}, demonstrating inconsistency of the existence of the $\Lambda nn$ bound state with $\Lambda N$ scattering as well as 3- and 4-body hypernuclear data. Furthermore, the renewed analysis of the BNL-AGS-E906 experiment \cite{BBGP19} led to conclusion that the formation of a bound $\Lambda nn$ nucleus is highly unlikely. In addition, rather recently Gal and Garzilazo \cite{GG19} made a rough but solid estimate of $\Lambda nn$ lifetime which, if bound, is considerably longer than the one of free $\Lambda$ hyperon $\tau_\Lambda$. This result is in disagreement with the shorter $\Lambda nn$ lifetime with respect to $\tau_\Lambda$ extracted from the HypHI events assigned to this system. The $\Lambda nn$ was also explored within pionless effective field theory (\nopieft) \cite{ARO15,HH19}.

In spite of the apparent interest the $\Lambda nn$ and $^3_\Lambda$H$^*$ continuum states have been investigated in only few theoretical works~\cite{BRS08,AG15,SBBM20}. Afnan and Gibson \cite{AG15} performed Faddeev calculations of $\Lambda nn$ using two-body separable potentials fitted to reproduce $NN$ and $\Lambda N$ scattering lengths and effective ranges. They pointed out that while $\Lambda nn$ pole appears in the subthreshold region (Re($E$)$<$0), only small increase of the $\Lambda N$ interaction strength produces a $\Lambda nn$ resonance (Re($E$)$>$0). This work encouraged the search for the $\Lambda nn$ system in the JLab E12-17-003 experiment \cite{JlabE17003}.

In this work, we performed few-body calculations of the $\Lambda nn$
and ${\rm ^3_\Lambda H^*}$ hypernuclear systems within LO \nopieft, both in the bound and continuous region, exploring thoroughly their nature. The first selected results have been reported in Ref. \cite{SBBM20}. As demonstrated in that work the virtual state ${\rm ^3_\Lambda H^*}$ pole position close to the $\Lambda+d$ threshold strongly affects the $\Lambda d$ $s$-wave phase shifts in $J^\pi=3/2^+$ channel. The calculated $\Lambda d$ scattering lengths and effective ranges from this work were further employed by Haidenbauer in the study of $\Lambda d$ correlation functions within the Lednicky-Lyuboshits formalism \cite{haidenbauer20}. It is to be noted that the nature of the ${\rm ^3_\Lambda H^*}$ state  is a subject of the JLab proposal P12-19-002 \cite{JlabP19002}.

The \nopieft approach was applied to $s$-shell $\Lambda$ hypernuclei and, among others, the experimental value of the $\Lambda$ separation energy $B_{\Lambda}$ in $^5_{\Lambda}$He was successfully reproduced~\cite{CBG18}. The \nopieft was further extended to $S=-2$ sector with the aim to study the onset of binding in $\Lambda \Lambda$ hypernuclei~\cite{CSBGM19}. 
Finally, in the present work the \nopieft is applied to the study of continuum states in 3-body hypernuclear systems. Bound state calculations are performed using the Stochastic Variational Method (SVM), the continuum states are described within the Inverse Analytic Continuation in the Coupling Constant (IACCC) Method and the Complex Scaling Method (CSM). The IACCC calculations are benchmarked against the CSM and the stability of resonance solutions is discussed. The CSM is in addition used to set limits of the accuracy of performed calculations. 

The paper is organized as follows: In Section II, we first give a brief description of the \nopieft approach and the SVM method applied in the calculations of few-body hypernuclear systems.  
Then, we introduce the CSM and IACCC method used to describe continuum states and pole movement in a complex energy plane.  
In Section III, we present results of our study of the $\Lambda nn$ and ${\rm ^3_\Lambda H^*}$ systems. We discuss in more detail the relation between the applied LO \nopieft approach and phenomenological models and, in particular, 
the stability and numerical accuracy of our \nopieft calculations. Finally, we summarize our findings in Section IV.

\section{Model and Methodology}
Hypernuclear systems studied in this work are described within the \nopieft at LO which was introduced in detail in \cite{CBG18}. In this section we present only basic ingredients of the theory. The LO \nopieft contains 2- and 3-body $s$-wave contact interaction terms, each of them associated with corresponding isospin-spin channel. The contact terms are then regularized by applying a Gaussian regulator with momentum cutoff $\lambda$. This procedure yields two-body $V_2$ and three-body $V_3$ potentials which together with the kinetic energy $T_{\rm k}$ enter the total Hamiltonian $H$ : 
\begin{equation} \label{efthamilt}
H=T_{\rm k}+V_2+V_3,
\end{equation}
where 
\begin{equation}
V_2 =\sum_{I,S} C_\lambda^{I,S} \sum_{i<j}\mathcal{P}^{I,S}_{ij} {\rm e}^{-\frac{\lambda^2}{4}r_{ij}^2} 
\end{equation}
and 
\begin{equation}
V_3 =\sum_{I,S} D_\lambda^{I,S} \sum_{i<j<k}\mathcal{Q}^{I,S}_{ijk}\sum_{\rm cyc}{\rm e}^{-\frac{\lambda^2}{4}\left(r_{ij}^2+r_{jk}^2\right)}.
\end{equation}
Here, $\mathcal{P}_{ij}^{I,S}$ and $\mathcal{Q}_{ijk}^{I,S}$ are the projection operators to 2- and 3-body $s$-wave isospin-spin $(I,S)$ channels and the 2- and 3-body low energy constants (LECs) $C_\lambda^{I,S}$ and $D_\lambda^{I,S}$ are fixed for each $\lambda$ by experimental data. The momentum cutoff $\lambda$ might be understood as a scale parameter with respect to a typical momentum $Q$. Calculated observables exhibit residual cutoff dependence $\mathcal{O}(Q/\lambda)$ suppressed with $\lambda$ approaching the renormalization group invariant limit $\lambda \rightarrow \infty$ \cite{CBG18}.

There are in total 4 two-body ($NN$, $\Lambda N$) and 4 three-body ($NNN$, $\Lambda NN$) LECs. Nuclear LECs $C_\lambda^{I=0,S=1}$, $C_\lambda^{I=1,S=0}$, and $D_\lambda^{I=1/2,S=1/2}$ are fitted to the deuteron binding energy,  $NN$ spin-singlet scattering length $a_0^{NN}$, and to the triton binding energy, respectively. Hypernuclear two-body LECs $C_\lambda^{I=1/2,S=0}$ and $C_\lambda^{I=1/2,S=1}$ are fixed by the $\Lambda N$ scattering length in a spin-singlet $a_0^{\Lambda N}$ and spin-triplet $a_1^{\Lambda N}$ channel.   Three-body hypernuclear LECs $D_\lambda^{I=0,S=1/2}$, $D_\lambda^{I=1,S=1/2}$, and $D_\lambda^{I=0,S=3/2}$ are fitted to the experimental values of $\Lambda$ separation energies $B_\Lambda ({\rm ^3_\Lambda H})$, $B_\Lambda ({\rm ^4_\Lambda H})$ 
and the excitation energy $E_{\rm exc} ({\rm ^4_\Lambda H^*})$. Here, we consider only $N$ and $\Lambda$ degrees of freedom, however, the effect of   
 the $\Lambda$-$\Sigma$ conversion process is implicitly accounted for by the chosen hypernuclear contact interaction. On the two-body level, we fit LEC to different values of $\Lambda N$ scattering lengths which represent strength of the free-space $\Lambda N$ interaction containing beside $\Lambda N - \Lambda N$ part also $\Lambda N -\Sigma N- \Lambda N$ contribution. On the few-body level the $\Lambda$-$\Sigma$ conversion is partially included in the three-body $\Lambda NN$ contact terms which are fixed using experimental values of the $\Lambda$ separation energies in 3- and 4-body hypernuclear systems.

 Since $a_0^{\Lambda N}$ and $a_1^{\Lambda N}$ are not constrained sufficiently well by experiment, we use their values given by direct analysis of scattering data \cite{AKS68} or predicted by several models of $\Lambda N$ interaction \cite{RSY99,PHM06,HPKM13}. Considered $a_0^{\Lambda N}$ and $a_1^{\Lambda N}$ together with the data used to fix $NN$ spin-singlet $^1S_0$ and spin-triplet $^3S_1$ LECs are given in Table.~\ref{tab1}.
 The \nopieft approach was applied to s-shell $\Lambda$ hypernuclei and, among others, the experimental value of the $\Lambda$ separation energy $B_{\Lambda}$ in $^5_{\Lambda}$He was successfully reproduced~\cite{CBG18} as demonstrated in the last column of Table~\ref{tab1}.

\begin{table}
\caption{\label{tab1} 
Values of spin-singlet $a^{\Lambda N}_0$ and spin-triplet $a^{\Lambda N}_1$ scattering lengths\footnote{We use the effective range expansion sign convention defined as $k{\rm cotg}(\delta)=-\frac{1}{a_s}+\frac{1}{2}r_s k^2+\cdots$ .} used to fit hypernuclear 2-body LECs together with effective ranges $r^{\Lambda N}_0$ and $r^{\Lambda N}_1$ (in fm). 
Corresponding $\Lambda$ separation energies $B_{\Lambda}({\rm ^5_\Lambda He};\infty)$ (in MeV), predicted within \nopieft for $\lambda \rightarrow \infty$ \cite{CBG18} are to be compared with the experimental value $B_{\Lambda}({\rm ^5_\Lambda He})=3.12(2)$~MeV \cite{davis05}.
}
\begin{ruledtabular}
\begin{tabular}{lccccc}
&$a^{\Lambda N}_0$&$r^{\Lambda N}_0$&$a^{\Lambda N}_1$&$r^{\Lambda N}_1$& $B_{\Lambda}({\rm ^5_\Lambda He}; \infty)$ \\ \hline
Alexander B \cite{AKS68}&-1.80&2.80&-1.60&3.30&3.01(10)\\
NSC97f \cite{RSY99}&-2.60&3.05&-1.71&3.33&2.74(11)\\
$\chi$EFT(LO) \cite{PHM06}&-1.91&1.40&-1.23&2.20&3.96(08)\\
$\chi$EFT(NLO) \cite{HPKM13}&-2.91&2.78&-1.54&2.27&3.01(06)\\ \hline
$NN$ \cite{MNS90,LA82} &-18.63 &2.75&\multicolumn{3}{c}{$E_{\rm B}({\rm ^2H})=-2.22457$ MeV}\\
\end{tabular}
\end{ruledtabular}
\end{table}

The calculation of $A=3,4,5$ -body $s$-shell $\Lambda$ hypernuclear systems is performed within finite basis set of correlated Gaussians \cite{VS95}
\begin{equation}
    \psi_i = {\hat \mathcal{A}}~{\rm exp}\left(-\frac{1}{2} {\bf x}^T A_i {\bf x}\right) \chi^i_{S M_S} \xi^i_{I M_I},
    \label{corrgauss}
\end{equation}                                               
where the operator $\hat \mathcal{A}$ ensures antisymmetrization between nucleons, ${\bf x}^T=({\bf x}_1,\dots,{\bf x}_{A-1})$ is a set of Jacobi coordinates, and $\chi^i_{S M_S}$ and $\xi^i_{I M_I}$ stand for corresponding spin and isospin parts, respectively. Each $\psi_i$ includes $A(A-1)/2$ nonlinear parameters which are placed in the $(A-1)$ dimensional positive-definite symmetric matrix $A_i$ plus 2 discrete parameters which represent different spin and isospin configuration in $\chi^i_{S M_S}$ and $\xi^i_{I M_I}$, respectively.

In order to choose $\psi_i$ with the most appropriate nonlinear parameters we use the Stochastic Variational Method (SVM) \cite{SV98} which was proved to provide systematic procedure to optimize the finite basis set, thus reaching highly accurate bound state description.

Resonances and virtual states, predominantly interpreted as poles of $S$-matrix \cite{taylor72,KKH89}, can not be addressed directly using the SVM with the finite basis set. Consequently, in order to study hypernuclear continuum we apply the Inverse Analytic Continuation in the Coupling Constant (IACCC) method \cite{HP17} which was proposed as numerically more stable alternative to the Analytic Continuation in the Coupling Constant \cite{accc77-79}.

Following the spirit of analytical continuation techniques we supplement the Hamiltonian $H$ (\ref{efthamilt}) by an auxiliary 3-body attractive potential
\begin{equation}
 V^{\rm IACCC}_3 = d_\lambda^{I,S} \sum_{i<j<k}\mathcal{Q}^{I,S}_{ijk}\sum_{\rm cyc}{\rm e}^{-\frac{\lambda^2}{4}\left(r_{ij}^2+r_{jk}^2\right)},
 \label{v3iaccc}
\end{equation}
where the amplitude $d_\lambda^{I,S}$ defines its strength and is negative for attraction. The projection operator $\mathcal{Q}_{ijk}^{I,S}$ ensures that the potential affects only a particular $(I,S)$ three-body channel - $(1,\frac{1}{2})$ for $\Lambda nn$ or $(0,\frac{3}{2})$ for $\rm ^3_\Lambda H^*$. If not explicitly mentioned, $\lambda$ in $V^{\rm IACCC}_3$ is equal to the \nopieft cutoff $\lambda$. In principle one can use a rather large class of 2- or 3-body attractive auxiliary potentials which fulfill certain criteria imposed by analytic continuation \cite{KKH89}. 
Using $V^{\rm IACCC}_3$ (\ref{v3iaccc}) ensures that the properties of 2-body part of the \nopieft Hamiltonian (\ref{efthamilt}) such as scattering lengths or deuteron binding energy remain unaffected. Its form is selected to be the same as of the \nopieft 3-body potential~(\ref{efthamilt}).

With increasing attractive strength of $d_\lambda^{I,S}$ the resonance or virtual state $S$-matrix pole described by $H$ starts to move towards the bound state region and at certain $d_{0,~\lambda}^{I,S}$ becomes a bound state. The other way around, studying bound state energy $E_{\rm B}$ as a function of $d_\lambda^{I,S}<d_{0,~\lambda}^{I,S}$ we can perform an analytic continuation of the pole position from the bound region back into the continuum ($d_\lambda^{I,S}>d_{0,~\lambda}^{I,S}$) up to the point of its physical position with no auxiliary force ($d_\lambda^{I,S}=0$).

In practice, we apply the SVM to calculate a set of $M+N+1$ bound state energies for different values of the coupling constant $\{E_{\rm B}^i(d_i);~d_i<d_0;~i=1,\dots,M+N+1\}$, where $d_i=d_{i,~\lambda}^{I,S}$. Next, using this set we construct the Pad\'{e} approximant of degree ($M$,$N$) $\mathcal{P}^{(M,N)}$ of function $d(\kappa)$
\begin{equation}
    \mathcal{P}^{(M,N)}(\kappa) =\frac{\sum_{j=0}^M b_j \kappa^j}{1+\sum_{j=1}^N c_j \kappa^j} \approx d(\kappa),
    \label{padeapprox}
\end{equation}
where $b_j$ and $c_j$ are real parameters of the $\mathcal{P}^{(M,N)}$. The $\kappa$ is defined as $\kappa=-{\rm i}k=-{\rm i}\sqrt{E}$ with $E$ standing for a bound state energy with respect to the nearest dissociation threshold. The position of the $S$-matrix pole corresponding to $H$ is calculated setting $d=0$ in Eq.~(\ref{padeapprox}) which leads to the the simple polynomial equation 
\begin{equation}
    \sum_{j=0}^M b_j \kappa^j=0.
    \label{poleq}
\end{equation}
The resonance or virtual state energy with respect to the nearest threshold is then obtained as $E=({\rm i}\kappa)^2$, where $\kappa$ now corresponds to the physical root of Eq.~(\ref{poleq}). Here, for complex resonance energy, we use the notation $E=E_r~-~ {\rm i}\Gamma/2$, where $E_r={\rm Re}(E)$ is the position of the resonance and $\Gamma=-2~{\rm Im}(E)$ stands for the resonance width.

Using the IACCC method we study the whole pole trajectory $E(d)$ in the continuum region $d \in \left< d_0;0 \right>$ (see Fig.~\ref{fig4}). For a given set of bound state energies $\{E_{\rm B}^i(d_i);~d_i<d_0;~i=1,\dots,M+N+1\}$, we shift $d_i \rightarrow d-d_i$ in the $E_{\rm B}^i(d_i)$ set, construct new Pad\'{e} approximant (\ref{padeapprox}), and obtain $E(d)$ as a corresponding root of Eq.~(\ref{poleq}).

The specific choice of $V^{\rm IACCC}_3$ (\ref{v3iaccc}) provides clear physical interpretation for any $d_\lambda^{I,S}$ solution. By varying $d_\lambda^{I,S}$ the $\Lambda nn$ or $\rm ^3_\Lambda H^*$ pole moves along its trajectory $E(d_\lambda^{I,S},\lambda)$ which is defined purely by the underlying 2-body interactions and cutoff $\lambda$. Supplementing the physical Hamiltonian (\ref{efthamilt}) by $V^{\rm IACCC}_3$ might be understood as a shift of the three-body LEC constant $D_\lambda^{I,S} \rightarrow D_\lambda^{I,S}+d_\lambda^{I,S}$. Since $D_\lambda^{I=1,S=1/2}$ and $D_\lambda^{I=0,S=3/2}$ have been fitted for each $\lambda$ to the experimental value of $B_\Lambda({\rm ^4_\Lambda H})$ and $E_{\rm exc}({\rm ^4_\Lambda H^*})$, respectively \cite{CBG18}, one could assign the parts of trajectories for $d_\lambda^{I,S}<0$ to an overbound 4-body system. In other words, for a given set of $a_0^{\Lambda N}$ and cutoff $\lambda$ the trajectory $E(d_\lambda^{I=1,S=1/2},\lambda)$ of $\rm \Lambda nn$ pole positions corresponds to different values of $B_\Lambda({\rm ^4_\Lambda H})$ and similarly the trajectory $E(d_\lambda^{I=0,S=3/2},\lambda)$ of $\rm ^3_\Lambda H^*$ pole positions corresponds to different values of $E_{\rm exc}({\rm ^4_\Lambda H^*})$.


For each IACCC resonance calculation we benchmark part of the corresponding pole trajectory against the Complex Scaling Method (CSM) \cite{AC71}.  
The main ingredient of the CSM is a transformation $U(\theta)$ of relative coordinates $\bf{r}$ and their conjugate momenta $\bf{k}$
\begin{equation}
U(\theta){\bf r} = {\bf r}e^{{\rm i}\theta},~~~~U(\theta){\bf k} = {\bf k}e^{-{\rm i}\theta},
\label{csmtrans}
\end{equation}
where $\theta$ is a real positive scaling angle. Applying this transformation to the Schr\"{o}dinger equation one obtains its complex scaled version
\begin{equation}
    H(\theta)\Psi(\theta)=E(\theta)\Psi(\theta),
    \label{csmsch}
\end{equation}
where $H(\theta)=U(\theta)HU^{-1}(\theta)$ is the complex scaled Hamiltonian and $\Psi(\theta)=U(\theta)\Psi$ is the corresponding wave function. For large enough $\theta$, the divergent asymptotic part of the resonance wave function is suppressed and $\Psi(\theta)$ is normalizable - possible resonant states can then be obtained as discrete solutions of Eq.~(\ref{csmsch}) \cite{AMKI06}. In order to prevent divergence of the complex scaled Gaussian potential (\ref{efthamilt}) the scaling angle is limited to $\theta < \frac{\pi}{4}$.

A mathematically rigorous formulation of the CSM for a two-body system results in the ABC theorem \cite{AC71} which provides description of the behavior of a complex scaled energy $E(\theta)$ with respect to $\theta$: (i) Bound state energies remain unaffected (ii) The continuum spectrum rotates clockwise in a complex energy plane by angle $2\theta$ from the real axis with its center of rotation at the corresponding threshold (iii) For $\theta>\theta_r=\frac{1}{2}{\rm arctan}\left(\frac{\Gamma}{2 E_r}\right)$ corresponding to the resonance energy $E_r$ and width $\Gamma$, the resonance is described by a square-integrable function and its energy and width are given by a complex energy $E(\theta)=E_r-{\rm i}\Gamma/2$ which does not change further with increasing $\theta$.

In this work, we expand $\Psi(\theta)$ in a finite basis of correlated Gaussians (\ref{corrgauss})
\begin{equation}
    \Psi(\theta)=\sum_{i=1}^N c_i(\theta)~\psi_i.
    \label{csmwf}
\end{equation}
Both resonance energies $E(\theta)$ and corresponding coefficients $c_i(\theta)$ are then obtained using the $c$-variational principle \cite{MCW78} as a solution of generalized eigenvalue problem 
\begin{equation}
\sum_{j=1}^N \left(\psi_i|H(\theta)|\psi_j \right)c_j^\alpha(\theta)=E^{\alpha}(\theta)\sum_{j=1}^N \left(\psi_i|\psi_j \right) c^{\alpha}_j(\theta),
\label{cgep}
\end{equation}
where $(|)$ stands for the $c$-product (bi-orthogonal product) \cite{AMKI06,moiseyev11}. In the case of real $\psi_i$, the c-product in Eq.~(\ref{cgep}) is equivalent to the inner product $<|>$. It was proved that the solutions of Eq.~(\ref{cgep}) are stationary in the complex variational space, and for $N \rightarrow \infty$ they are equal to exact solutions of the complex scaled Schr\"{o}dinger equation (\ref{csmsch}) \cite{MCW78}. Nevertheless, with increasing number of basis states the solution stabilizes and there is no upper or lower bound to an exact resonance solution \cite{moiseyev98}.  

In fact, due to a finite dimension of the basis set the resonance energy $E(\theta)$ (\ref{cgep}) moves with increasing scaling angle along the $\theta$-trajectory even for $\theta > \theta_r$, featuring residual $\theta$ dependence \cite{AMKI06,MKV07}. It was demonstrated that following the generalized virial theorem \cite{MCW78,YW78} the best estimate of a resonance energy is given by the most stationary point of the $\theta$-trajectory, i.e. such $E(\theta_{\rm opt})$ for which the residual $\theta$ dependence is minimal but not necessarily equal to zero
\begin{equation}
    \left|\frac{{\rm d}E(\theta)}{{\rm d}\theta}\right|_{\theta_{\rm opt}}\approx 0.
\label{statcond}
\end{equation}

A real scaling angle $\theta$ is frequently used in finite basis CSM calculations with satisfactory results \cite{FGJ03,MKV07,HIKMM15}. However, identifying the resonance energy with $E(\theta_{\rm opt})$ using the $\theta$-trajectory (Im$(\theta)=0$, Re$(\theta)$ changing) is still approximate. As pointed out by Moiseyev \cite{moiseyev98} the resonance stationary condition requires exact equality in Eq.~(\ref{statcond}), which can be achieved in a finite basis set by considering complex $\theta_{\rm opt}$. Consequently, taking $\theta$ real introduces certain theoretical error and it is problematic to quantify how much the result obtained using $\theta$-trajectory technique deviates from the true CSM resonance solution (zero derivative in Eq.~(\ref{statcond})).

Following Aoyama et al. \cite{AMKI06} we use both $\theta$-trajectory and $\beta$-trajectories (Re$(\theta)$ fixed, Im$(\theta)$ changing) to locate the position of the true CSM solution. In the above work it was numerically demonstrated that for certain Re$(\theta_{\rm opt})$ the $\theta$-trajectory approaches the stationary point and then starts to move away. On the other hand, the $\beta$-trajectories are roughly circles with decreasing radius as the corresponding Re$(\theta)$ approaches Re$(\theta_{\rm opt})$. In view of orthogonality of the $\theta$- and $\beta$-trajectories at given scaling angle $\theta$, the true CSM solution is then located inside an area given by circular $\beta $-trajectories. More specifically, it is identified as the center of the circular $\beta $-trajectory with the smallest radius where the CSM error is given by the size of this radius \cite{AMKI06}. 
 
Another non-trivial task is to determine an appropriate yet not excessively large correlated Gaussian basis which yields stable CSM resonance solution. In this work, we apply the HO trap technique \cite{SBBM20} which introduces systematic algorithm how to select such basis. First, we place a resonant system described by the Hamiltonian $H$ into a harmonic oscillator (HO) trap
\begin{equation}
H^{\rm trap}(b)=H+V^{\rm HO}(b),~~~V^{\rm HO}(b)=\frac{\hbar^2}{2mb^4} \sum_{j<k}r_{jk}^2,
\label{hotrap}
\end{equation}
where $b$ is the HO trap length and $m$ is an arbitrary mass scale. Next, for given $b$ we apply the SVM to determine basis set which yields accurate description of the ground as well as  excited states of $H^{\rm trap}(b)$. The potential $V^{\rm HO}(b)$ plays a role analogous to a box boundary condition (though not so stringent) -- the SVM procedure promotes basis states with their typical radius given by the trap length $b$.
By increasing $b$ we enlarge the typical radius of the correlated Gaussians $\psi_i$. 
For large enough $b$, the CSM resonance solution for the Hamiltonian $H$ starts to stabilize and both the short range and  suppressed long range asymptotic parts of a resonance wave function are described sufficiently well.

For $\theta \geq \theta_r$ the CSM resonant wave function $\Psi(\theta)$ is localized in a certain interaction region whereas its asymptotic part is suppressed by the CSM transformation (\ref{csmtrans}). Consequently, we use the HO trap technique in order to build the CSM basis which describes physically relevant interaction region of $\Psi(\theta)$ up to certain large enough $R_{\rm max}$ beyond which the asymptotic part does not contribute significantly to the CSM solution .

In practice, for each CSM calculation, we apply the HO trap technique to independently select basis sets for a grid of increasing trap lengths $\{b_i; b_i \leq b_{\rm max}\}$. Next, merging these sets into a larger CSM basis we calculate the resonance $\theta$-trajectory solving Eq.~(\ref{cgep}). In the last step we study stabilization of the $\theta$-trajectory with increasing $b_{\rm max}$ considered in the merged basis set.  For more details and an example see Subsection~\ref{stability}.

\section{Results}
We applied the LO \nopieft approach with 2- and 3-body regulated contact terms defined in Eq.~(\ref{efthamilt}) to the study of the $s$-shell $\Lambda$ hypernuclei, the $\Lambda nn$ and ${\rm ^3_\Lambda H^*}(J^\pi=3/2^+,~I=0)$ systems in particular. In this section, we present results of the calculations and provide comparison of the results obtained within our LO \nopieft approach and phenomenological models. In a separate subsection, we discuss in detail stability and numerical accuracy of the presented SVM and IACCC resonance solutions. 

\begin{figure}
\includegraphics[width=0.5\textwidth]{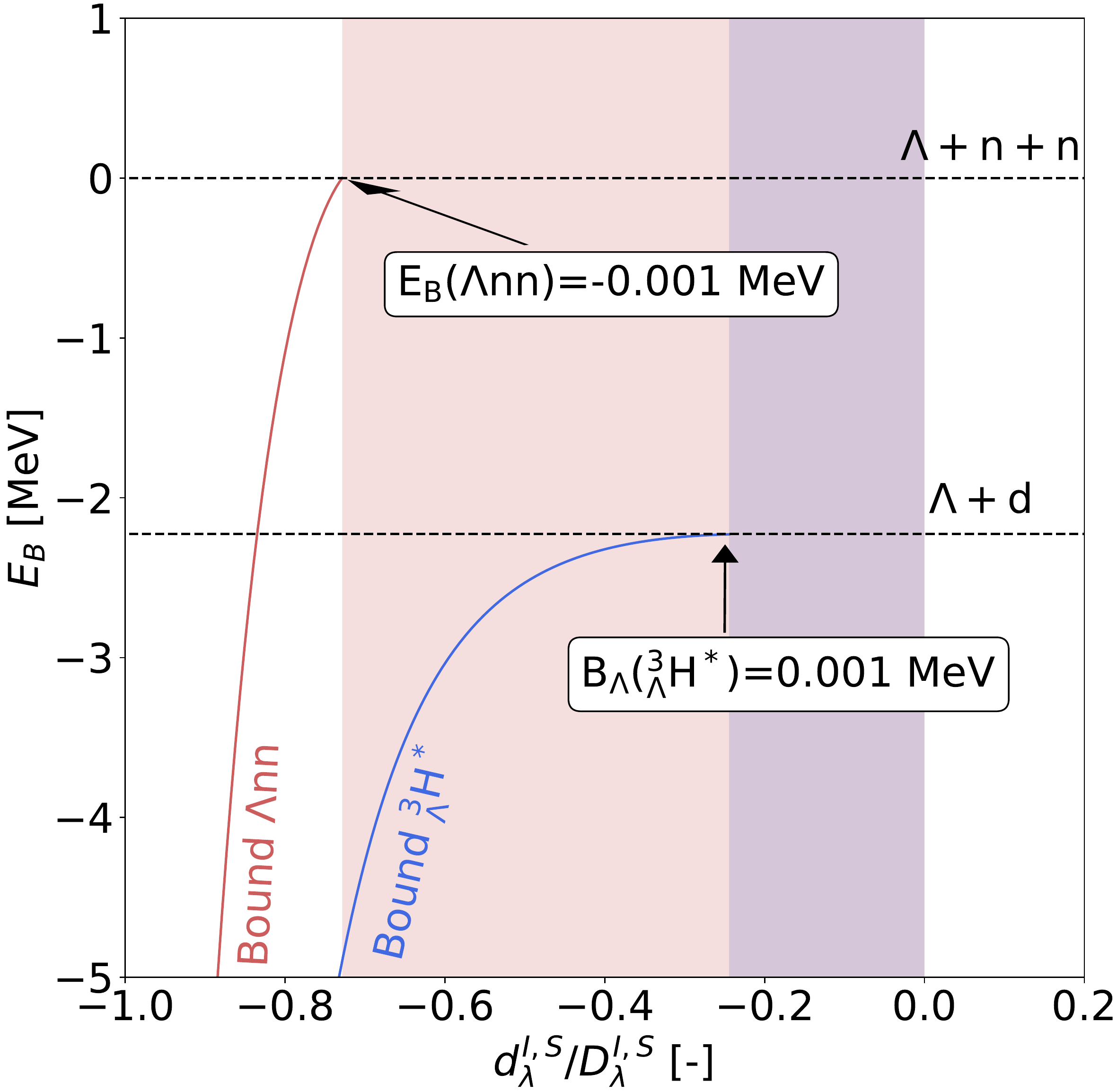}
\caption{\label{fig1} The $\Lambda nn$ and ${\rm ^3_\Lambda H^*}$ bound state energies $E_{\rm B}$ as a function of $d_\lambda^{I,S}$ normalized to $D_\lambda^{I,S}$ for $I=1, S=1/2$ and $I=0, S=3/2$, respectively. The calculation is performed for the Alexander B set of $\Lambda N$ scattering lengths and $\lambda=6~{\rm fm^{-1}}$.}
\end{figure}

The additional auxiliary 3-body potential $V^{\rm IACCC}_3$ (\ref{v3iaccc}) introduced to study continuum states allows us to vary the amount of attraction and thus explore different scenarios, as demonstrated in Fig.~\ref{fig1}. 
Here, the $\Lambda nn$ and $^3_{\Lambda}{\rm H}^*$ bound state energies $E_{\rm B}$ are plotted as a function of the strength $d^{I,S}_{\lambda}$ of the auxiliary force normalized to the strength $D^{I,S}_{\lambda}$ 
of the 3-body $\Lambda NN$ potential of the \nopieft. In the limiting case 
 $d^{I,S}_{\lambda}/D^{I,S}_{\lambda} =-1$, the 3-body repulsion is completely canceled and the systems undergo Thomas collapse \cite{thomas35} in the limit of $\lambda \rightarrow \infty$. For suitably chosen values of   $d^{I,S}_{\lambda}/D^{I,S}_{\lambda}$ between -1 and 0, both $\Lambda nn$ and $^3_{\Lambda}{\rm H}^*$ are bound and one can study implications for the 4- and 5-body 
 $s$-shell hypernuclei as will be shown below where we tune $d^{I,S}_{\lambda}$ to get either $\Lambda nn$ or $^3_{\Lambda}{\rm H}^*$ just bound by 0.001 MeV. 
 Finally, for the zero auxiliary force $d^{I,S}_{\lambda}/D^{I,S}_{\lambda} = 0$ one gets 
 physical solutions, namely continuum states of $\Lambda nn$ and $^3_{\Lambda}{\rm H}^*$ 
 (either resonant or virtual states). 
 The figure suggests that the value of $d^{I,S}_{\lambda}/D^{I,S}_{\lambda}$ considerably closer to 0, i.e. much less additional attraction, is needed to get $^3_{\Lambda}{\rm H}^*$ bound then in the case of $\Lambda nn$. 

\begin{figure*}[t!]
\includegraphics[width=0.9\textwidth]{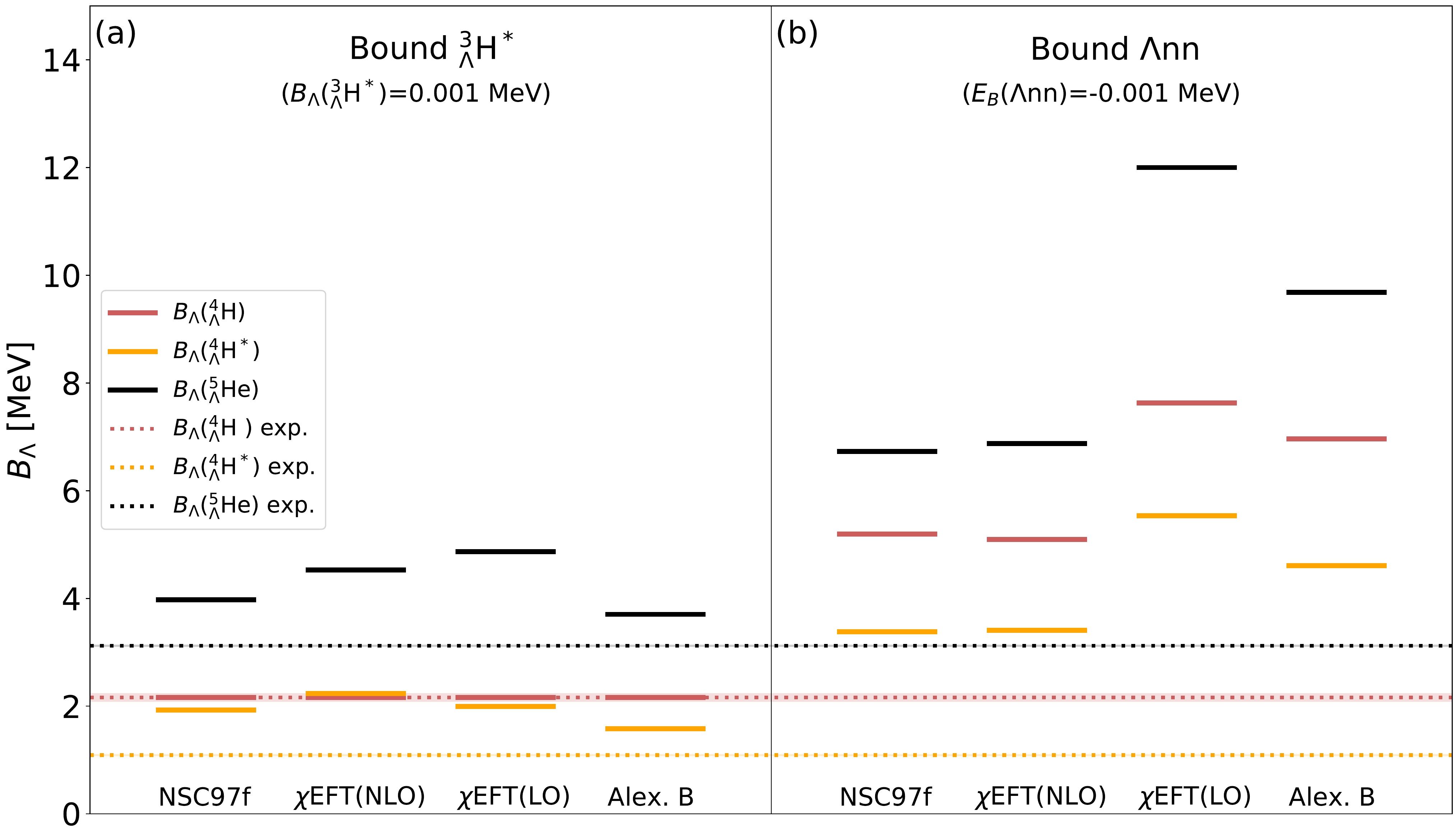}
\caption{\label{fig2}$\Lambda$ separation energies $B_\Lambda$ from SVM calculations using cutoff $\lambda=6~{\rm fm^{-1}}$ and several sets of $\Lambda N $ scattering lengths for two cases - just bound $\rm ^3_\Lambda H^*$ (a) and just bound $\Lambda nn$ (b). Horizontal dotted lines mark experimental values of $B_\Lambda$. }
\end{figure*}
We will now demonstrate that such $\Lambda$ interactions tuned to bind $\Lambda nn$ and/or  $\rm ^3_\Lambda H^*$ are inconsistent with $\Lambda$ separation energies in $A=4$ and 5 hypernuclei. We keep 2- and 3-body LECs fixed and fit the attractive strength of the auxiliary 3-body force, either $d_\lambda^{I=0,S=3/2}$ to $\Lambda$ separation energy $B_\Lambda({\rm ^3_\Lambda H^*})=0.001~{\rm MeV}$ or $d_\lambda^{I=1,S=1/2}$ to bound state energy $E_B({\Lambda nn})=-0.001~{\rm MeV}$.

Consequences of such tuning are illustrated in Fig.~\ref{fig2}. Here, we present $\Lambda$ separation energies $B_{\Lambda}$ in $s$-shell hypernuclei, calculated for selected $\Lambda N$ scattering lengths and cutoff $\lambda=6~{\rm fm^{-1}}$ which already exhibits partial renormalization group invariance. Variations of $d_\lambda^{I=0,S=3/2}$ or $d_\lambda^{I=1,S=1/2}$ do not affect the $I,S=\left(0,\frac{1}{2}\right)$ three-body channel, consequently, the $\Lambda$ separation energy of the hypertriton ground state remains unaffected and is not shown in the figure.
In order to get the $\rm ^3_\Lambda H^*$ system just bound (left panel (a)), the amount of repulsion in the $\left(0,\frac{3}{2}\right)$ three-body channel must decrease, which leads in return to overbinding of both the $\rm ^4_\Lambda H^*$ excited state and the $\rm ^5_\Lambda He$ hypernucleus. The wave function of the $\rm ^4_\Lambda H$ ground state does not include the $\left(0,\frac{3}{2}\right)$ component and thus its $B_\Lambda$ remains intact. 
As was already noted and demonstrated in Fig.~\ref{fig1}, 
the binding of the $\Lambda nn$ system requires a larger change in the corresponding auxiliary three-body force. Indeed, decreasing amount of repulsion in the $\left(1,\frac{1}{2}\right)$ three-body channel induces even more severe overbinding than in the $\rm ^3_\Lambda H^*$ case - $B_\Lambda$s are more than twice larger than experimental values (right panel (b)). 
We might deduce that by varying the strength of $\Lambda$ interactions, it is harder to get $\rm \Lambda nn$ bound - the bound $\rm ^3_\Lambda H^*$ state appears more likely first. This result is in agreement with previous works \cite{MKGS95,GFV07a,GFV07b}.\\ 

\begin{figure*}[t!]
\begin{tabular}{cc}
\includegraphics[width=0.45\textwidth]{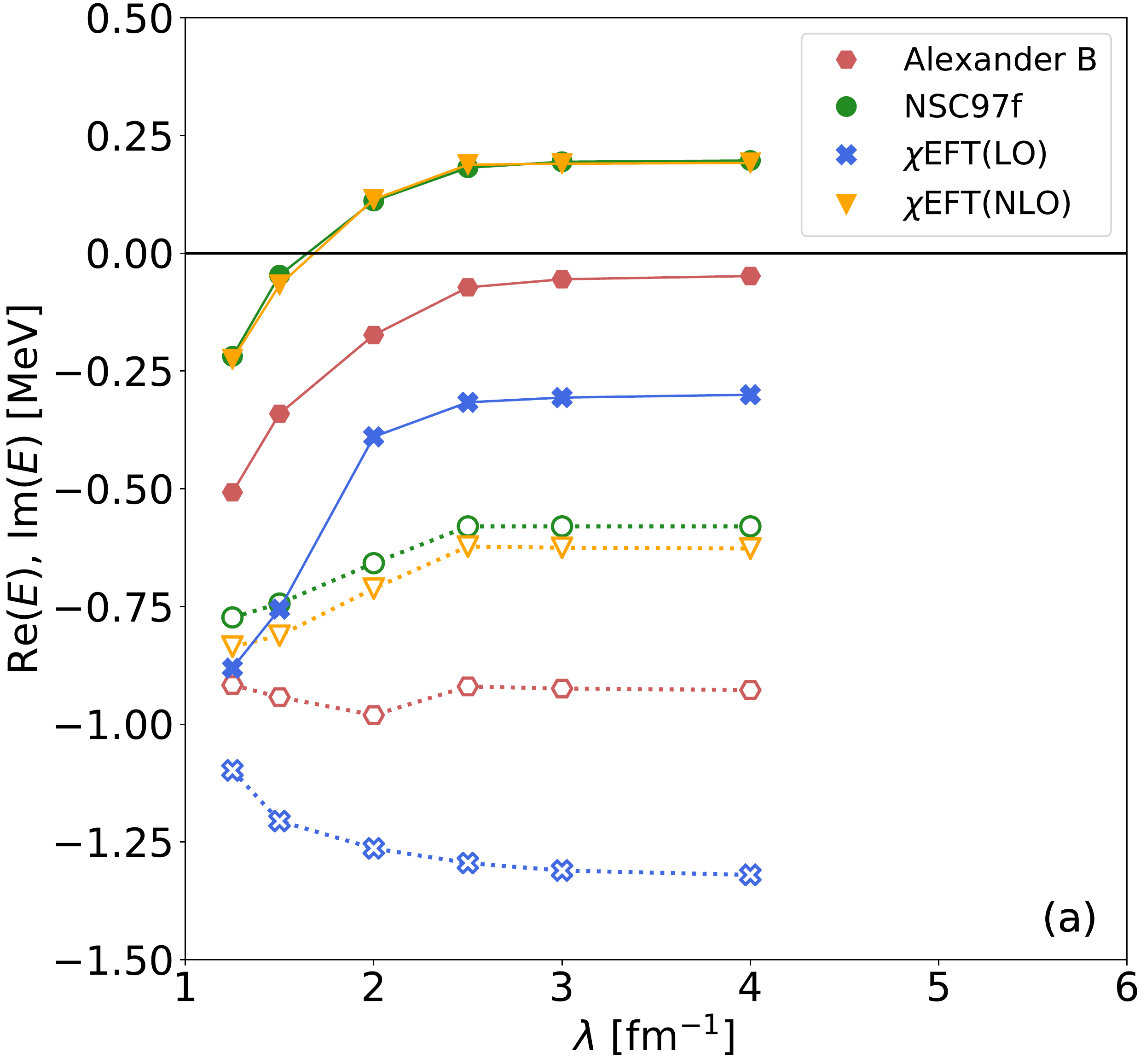}&\includegraphics[width=0.45\textwidth]{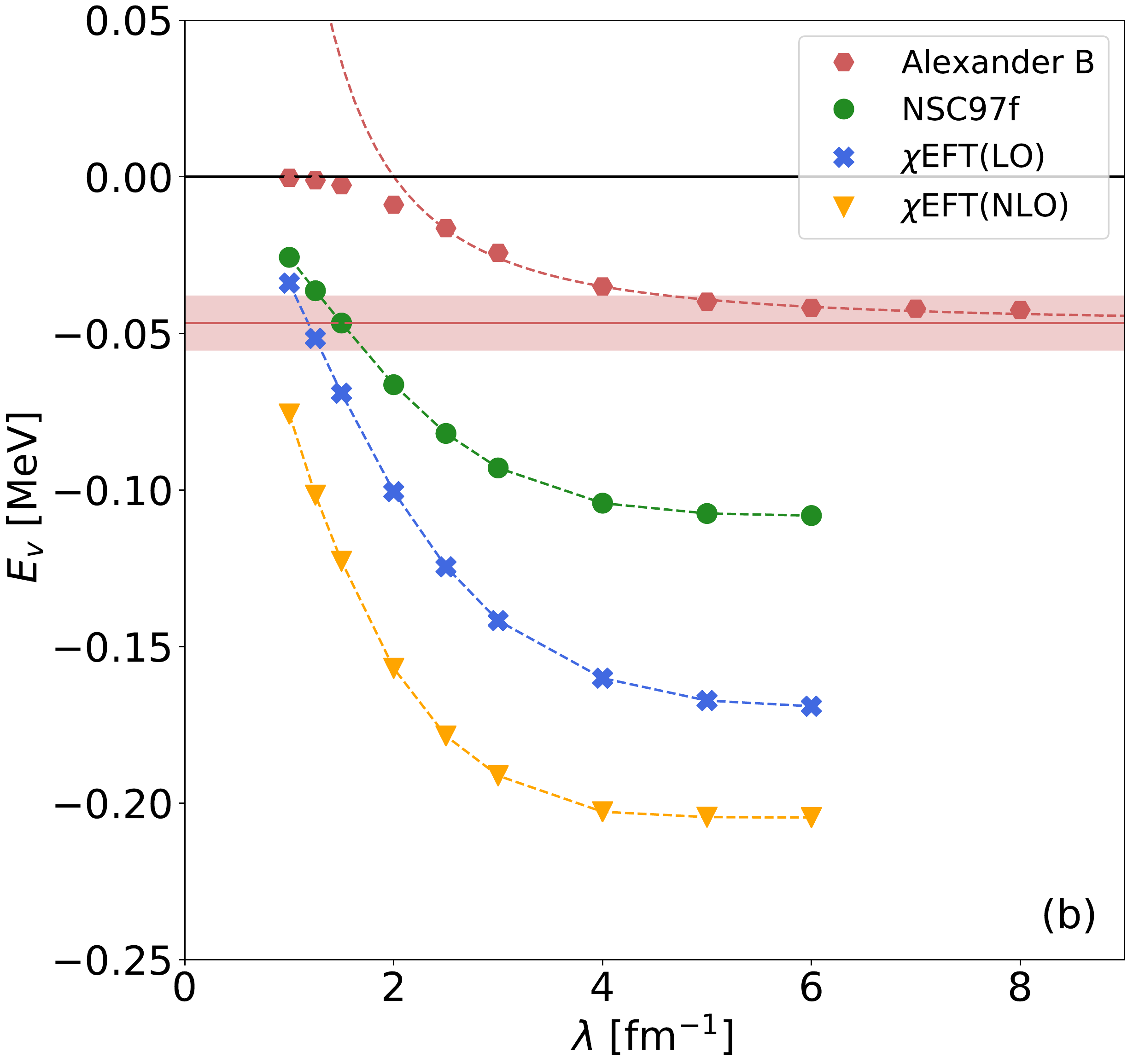}\\
\end{tabular}
\caption{\label{fig3}Real Re($E$) (full symbols) and imaginary Im($E$) (empty symbols) parts of the $\Lambda nn$ resonance energy (a) and energy $E_{\rm v}$ of the $\rm ^3_\Lambda H^*$ virtual state with respect to the $\Lambda+d$ threshold (b) as a function of cutoff $\lambda$ calculated using the IACCC method for several $\Lambda N$ interaction strengths. For $\rm ^3_\Lambda H^*$ virtual state and Alexander B we perform extrapolation for $\lambda \rightarrow \infty$. The red dashed line is the  extrapolation function, the solid red line and shaded area mark the contact limit and the extrapolation error. For theoretical error see the text, numerical errors are discussed in Subsection~\ref{stability}.}
\end{figure*}

In Fig.~\ref{fig3} we show the physical solutions (with no auxiliary force) corresponding to the \nopieft Hamiltonian $H$ (\ref{efthamilt}). Here, the real Re$(E)$ and imaginary Im$(E)$ parts of the $\Lambda nn$ resonance energy (left panel (a)) and the energy $E_{\rm v}$ of the virtual state $^3_{\Lambda}{\rm H}^*$ 
with respect to the $\Lambda+d$ threshold (right panel (b)) are plotted as a function of the cutoff $\lambda$ for the $\Lambda N$ scattering length versions listed in Table~\ref{tab1}. The calculated energies in the both hypernuclear systems depend strongly on the input $\Lambda N$ interaction strength. 
In the case of $^3_{\Lambda}{\rm H}^*$, we obtain for all considered $\Lambda N$ scattering lengths a virtual state solution. Namely, in accord with the definition of a virtual state~\cite{taylor72},  the imaginary part of the $^3_{\Lambda}{\rm H}^*$ pole momentum Im($k$) decreases from a positive value (bound state) to a negative value (unbound state) with a decreasing auxiliary attraction whereas the real part Re($k$) remains equal to zero~\cite{taylor72} (as was demonstrated in ref.~\cite{SBBM20}). 
On the other hand, in the case of the $\Lambda nn$ system the \nopieft predicts a resonant state. Moreover, only the NSC97f and $\chi$EFT(NLO) yield $\Lambda N$ interaction strong enough to ensure for $\lambda \ge 2$~fm$^{-1}$ the $\Lambda nn$ pole position in the fourth quadrant of a complex energy plane (Re$(E) > 0$, Im$(E) < 0$), i.e. predict a physical $\Lambda nn$ resonance.  

In Fig.~\ref{fig3} we also demonstrate stability of the solutions with respect to the cutoff $\lambda$. The calculated energies vary smoothly beyond the value $\lambda = 2$~fm$^{-1}$ and already at $\lambda = 4$~fm$^{-1}$ they stabilize within extrapolation uncertainties at an asymptotic value corresponding to the renormalization scale invariance limit  $\lambda \rightarrow \infty$. This is illustrated in the right panel, where we present for the Alexander B case the extrapolation function and the asymptotic value including the extrapolation error 
for the energy $E_{\rm v}$ of the $^3_{\Lambda}{\rm H}^*$ virtual state. 
It is to be noted that one might naively expect clear dependence on the strength of the  $\Lambda N$ spin-triplet interaction which solely enters the $\rm ^3_\Lambda H^*$ hypernuclear part on a two-body level. However, the dominance of the spin-triplet interaction is undermined by 3-body force in the $\left(0,\frac{3}{2}\right)$ channel compensating the size of the spin-singlet scattering length $a_0^{\Lambda N}$, being fixed by the $B_\Lambda({\rm ^4_\Lambda H^*})$ experimental value. 

One could argue that considering different values of $a^{\Lambda N}_s$ or strengths of $\Lambda NN$ three-body forces would open a possibility to locate the $ \Lambda nn$ resonance in the fourth quadrant closer to the real axis and thus decrease its width $\Gamma$. This would certainly facilitate its experimental observation. However, $\Lambda NN$ forces are fixed by experimental $B_\Lambda$s of 3- and 4-body hypernuclear systems. Considering unusually large values of $a^{\Lambda N}_s$ would allow $\Lambda nn$ pole position closer to the threshold but $\Lambda N$ interactions would have to be reconciled again with remaining $s$-shell systems. At LO \nopieft we would be constrained by a possibility of bound $^3_\Lambda \rm H^*$ and by the experimental value of $B_\Lambda({\rm ^5_\Lambda He})$.

In order to make an estimate of the effective range corrections in the 3-body hypernuclear systems, we consider that the relevant typical energy scale - i.e., the $\Lambda$ binding energy or the resonance energy - is small, and therefore it should be sensitive only to the long-distance properties of the $\Lambda N$ interaction \cite{hammer02}.
In our case, the relevant energies of both the $\Lambda nn$ resonance as well as the hypertriton virtual state are less than 1~MeV. For the hypertriton, one could estimate the typical $\Lambda$ momentum as $p_\Lambda \sim \sqrt{ 2\mu E} \approx 37~{\rm MeV}$, where $\mu \approx 700~{\rm MeV}$ is the $\Lambda-{\rm deuteron}$ reduced mass. The leading correction of the effective range should be of order $O(QR)$, where $Q\sim p_\Lambda$ and $R$ is the range of the $\Lambda$N interaction. This gives a truncation error of about $\approx 47\%$, where we consider $R \sim r_s^{\Lambda N} \approx 2.5~{\rm fm}$. The typical $\Lambda$  momentum in the $\Lambda nn$ resonance should be roughly the same.

Following Ref.~\cite{BEK16}, a rough estimate of the LO error can be made through residual cutoff dependence which has to be corrected by the NLO term. Inspecting the evolution of the $\Lambda nn$ and $\rm _\Lambda^3H^*$ energies plotted in Fig.~\ref{fig3} as a function of the cutoff from $\lambda=1.25$~${\rm fm}^{-1}\approx 250$~MeV to $\lambda \rightarrow \infty$ one can estimate the LO uncertainty. The residual cutoff dependence in Fig.~\ref{fig3} indeed gives estimation similar to the one based on the typical $\Lambda$ momentum. Moreover, the lowest cutoff in this plot represents calculations where the effective range is roughly reproduced, which can hint the NLO results. In any case, the truncation error is comparable with the uncertainty due to a different $a^{\Lambda N}_s$ input, and one can see that the calculated resonant and virtual state energies remain in the vicinity of the threshold.

\begin{figure*}[t!]
\includegraphics[width=0.9\textwidth]{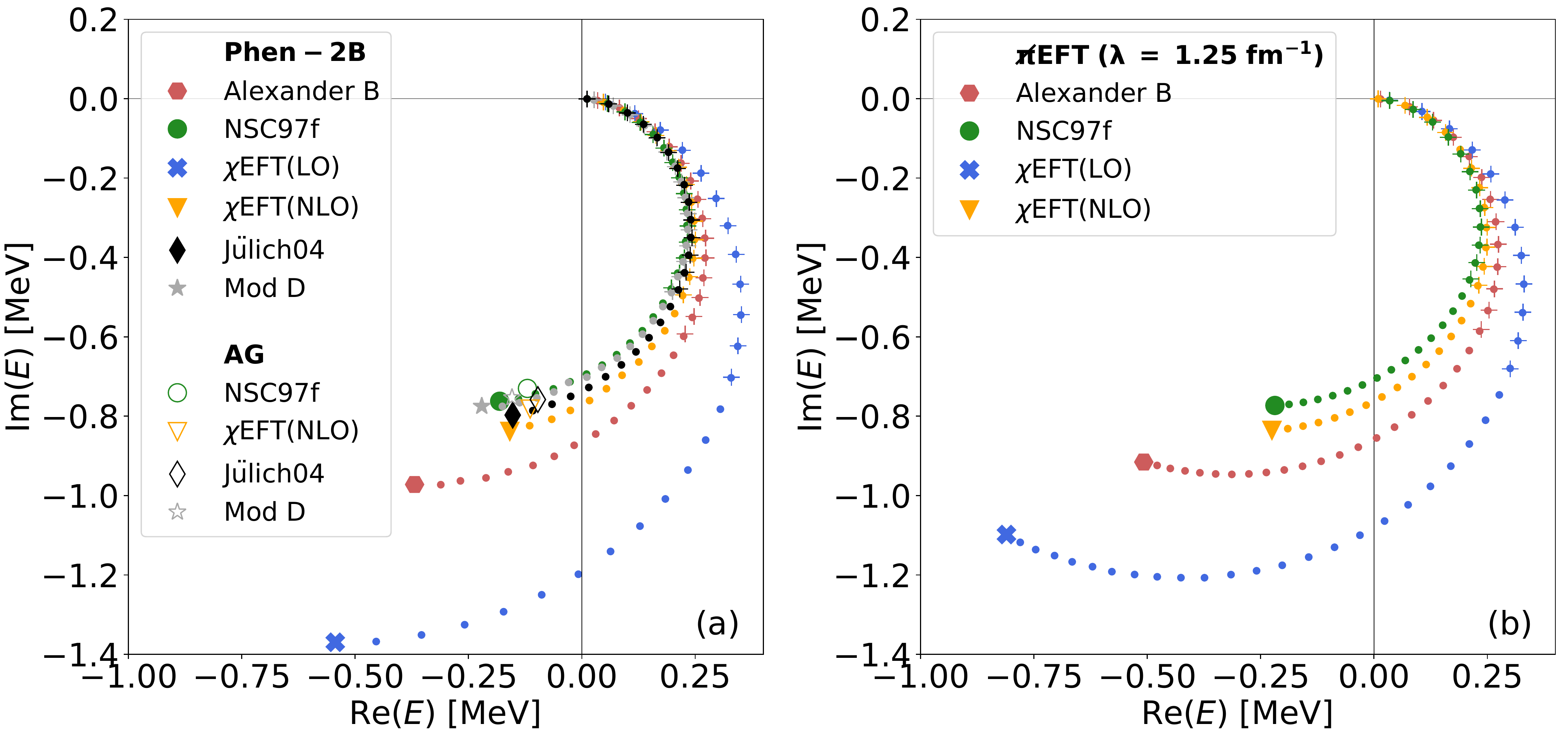}
\caption{\label{fig4}Trajectories of the $\Lambda nn$ resonance pole in a complex energy plane determined by a decreasing attractive strength $d_\lambda^{I=1,S=1/2}$ for several $\Lambda N$ interaction strengths. Left panel (a): calculations using $\Lambda N$ and $NN$ phenomenological potential Phen-2B (\ref{phen2B}). Larger full symbols stand for the physical position of the $ \Lambda nn$ pole ($d_\lambda^{I=1,S=1/2}=0$), empty symbols mark corresponding solutions obtained by Afnan and Gibson (AG)  \cite{AG15} for the same scattering lengths and effective ranges used to fix potential Phen-2B (\ref{phen2B}). 
Right panel (b): \nopieft calculations for cut-off $\lambda=1.25~{\rm fm^{-1}}$. 
In a region accessible by the CSM we also show for each IACCC solution (dots) the one obtained by the CSM (crosses) for the same amplitude of the auxiliary three-body force.}
\end{figure*}
Clearly, the issue of truncation error in \nopieft is not fully settled; see, for example, Ref.~\cite{griesshammer20}. A precise estimate of this error can be done only after calculating a few orders in the EFT expansion. To conclude, we dare to state that we do not expect the NLO effects to change qualitatively the LO results, i.e., the excited state of the hypertriton will remain a virtual state and the $\Lambda nn$ system will remain a resonance (either physical with ${\rm Re}(E) > 0$~MeV or unphysical with ${\rm Re}(E) < 0$~MeV). One can further speculate that since the $\Lambda nn$ resonance energy in Fig.~\ref{fig3} moves with decreasing cutoff (increasing induced effective ranges) into the third quadrant of a complex energy plane (Re$(E) < 0$, Im$(E) < 0$), inclusion of non-zero effective range through the NLO correction would more likely yield unphysical $\Lambda nn$ resonance.

Our work represents the first EFT study of the $\Lambda nn$ and $\rm ^3_\Lambda H^*$ hypernuclear systems in a continuum. Therefore, we find it appropriate to discuss difference of our approach with respect to the previous calculations of the $\Lambda nn$ resonance performed by Afnan and Gibson using a phenomenological approach \cite{AG15}. Following their work we neglect three-body force but instead of separable non-local two-body potentials we employ one range Gaussians

\begin{equation}
 V(r)=\sum_{I,S}\hat{\mathcal{P}}_{I,S}~C_{I,S}~{\rm exp}\left(-\frac{\lambda^2_{I,S}}{4}r^2\right)
 \label{phen2B}
\end{equation}
to describe $s$-wave interaction in nuclear $I,S=(0,1),~(1,0)$ and hypernuclear $I,S=(1/2,1),~(1/2,0)$ two-body channels. Here, $\hat{\mathcal{P}}_{I,S}$ is the projection operator. The parameters $C_{I,S}$ and $\lambda_{I,S}$ are fitted to the values of $a_s$ and $r_s$ listed in \cite{AG15}. Moreover, we took into account $a^{\Lambda N}_s$ and $r^{\Lambda N}_s$ related to Alexander B and $\chi$EFT(LO) given in Table~\ref{tab1}. 

The calculated $\Lambda nn$ pole trajectories for the Phen-2B potential (\ref{phen2B}) are presented in Fig.~\ref{fig4}, left panel (a). The auxiliary interaction is in a form of three-body force (\ref{v3iaccc}) with cutoff $\lambda=1~{\rm fm^{-1}}$. We observe that calculated physical pole positions (filled larger symbols) are in good agreement with those presented in \cite{AG15} (empty symbols). Indeed, as might be expected the position of the near-threshold $\rm \Lambda nn$ resonance is predominantly given by low-momentum characteristics of an interaction - $a_s$ and $r_s$ which are the same in both cases. 

In order to reveal the relation between the LO \nopieft and phenomenological approaches discussed above, one can consider the finite cutoff $\lambda_s$ which gives roughly the same values of $r_s$ as used in the above phenomenological calculations. Such a value, $\lambda_s \approx 1.25~{\rm fm^{-1}}$ for NSC97f and $\chi$EFT(NLO), yields in addition $B_\Lambda({\rm ^5_\Lambda He})$ remarkably close to experiment~\cite{CBG19}. As explained by the authors one might understand that $\lambda_s$ absorbs into LECs NLO contributions of the theory which are likely to increase its precision, however, success of this procedure is not in general guaranteed for all systems. Indeed, higher orders above NLO which behave as powers of $(Q/\lambda)$ are induced as well and are not suppressed by $\lambda \rightarrow \infty$. In Fig.~\ref{fig4}, right panel (b), we present  $\Lambda nn$ pole trajectories calculated using the \nopieft for this specific $\lambda_s$ value and several $\Lambda N$ interaction strengths. One notices very close positions of the $\Lambda nn$ resonance calculated for $\chi$EFT(NLO) and NSC97f using the Phen-2B potential (left panel (a)) and the \nopieft (right panel (b)). The LO \nopieft for $\lambda=1.25~{\rm fm^{-1}}$ could thus be considered as a suitable phenomenological model which yields good predictions for 4- and 5- body hypernuclei and hypertriton~\cite{CBG18,CBG19}. 

In addition, in both panels of Fig.~\ref{fig4} we compare the $\Lambda nn$ pole positions calculated within the CSM and IACCC method for the same values of $d_\lambda^{I=1,S=1/2}$ located in the area reachable by the CSM. We might see  remarkable agreement between IACCC (dots) and CSM (crosses) solutions, which provides benchmark of the calculations and demonstrates high precision of our results. 

\begin{figure*}
\includegraphics[width=0.9\textwidth]{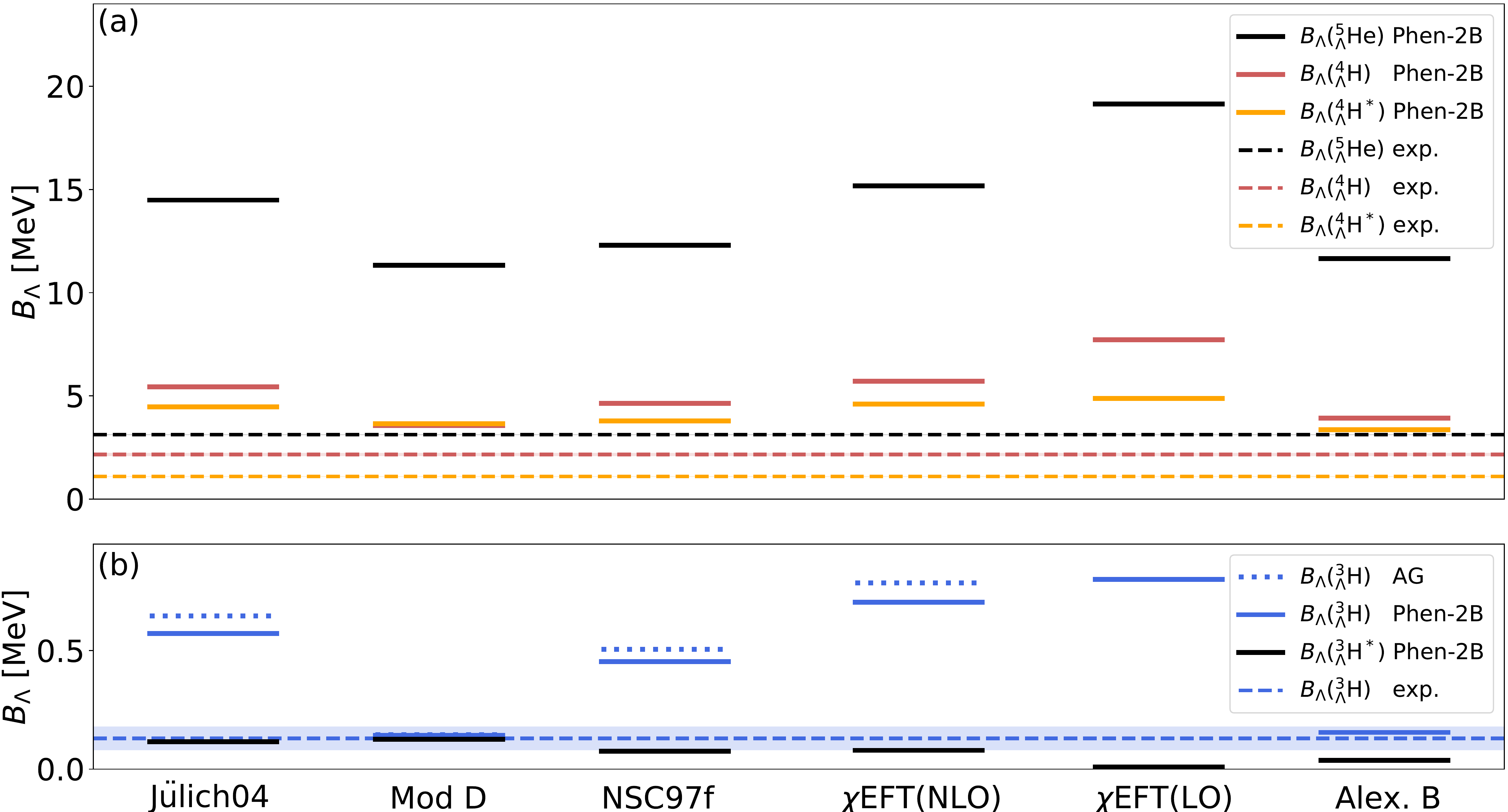}
\caption{\label{fig5} 
Upper panel (a): $\Lambda$ separation energies $B_\Lambda({\rm ^4_\Lambda H})$, $B_\Lambda({\rm ^4_\Lambda H^*})$, and $B_\Lambda({\rm ^5_\Lambda He})$ from SVM calculations using various $\Lambda N$ interaction strengths of the Phen-2B interaction (\ref{phen2B}). The nuclear part is given by the same form of a phenomenological potential. Experimental values of $B_\Lambda$ are marked by dashed horizontal lines. Lower panel (b): the same for $B_\Lambda({\rm ^3_\Lambda H})$ and $B_\Lambda({\rm ^3_\Lambda H^*})$. Dotted lines show $B_\Lambda({\rm ^3_\Lambda H})$ obtained by Afnan and Gibson \cite{AG15}.}
\end{figure*}

In Fig.~\ref{fig5}, we show $B_\Lambda$ of remaining $s$-shell hypernuclear systems, calculated using the Phen-2B potential~(\ref{phen2B}). The hypertriton ground state $\rm ^3_\Lambda H$ is in most cases overbound, calculated $B_\Lambda ({\rm ^3_\Lambda H})$ are consistent with those obtained by Afnan and Gibson using separable non-local potentials fitted to the same $\Lambda N$ interaction strengths~\cite{AG15}. The excited state of  hypertriton $\rm ^3_\Lambda H^*$ turns to be bound, which is in disagreement with previous theoretical calculations  \cite{MKGS95,HOGR14}. Heavier $s$-shell systems are considerably overbound as well, regardless of which specific set of $a^{\Lambda N}_s$ and $r^{\Lambda N}_s$ is fitted.
Overbinding of $s$-shell hypernuclear systems brought about by the Phen-2B interaction (\ref{phen2B}) clearly indicates a missing piece which would introduce necessary repulsion. This could be provided by introducing a $\Lambda NN$ three-body force. In fact, Afnan and Gibson stated that more detailed study of the $\Lambda nn$ resonance including three-body forces should be considered \cite{AG15}. In \nopieft additional repulsion is included right through the $\Lambda NN$ force fitted for each cutoff $\lambda$ to experimental values of $B_\Lambda$ in 3- and 4-body hypernuclei. As a result, though both the Phen-2B (as well as AG) interaction  and the \nopieft for $\lambda = 1.25$~fm$^{-1}$ yield close positions of the $\Lambda nn$ resonance 
(see Fig.~\ref{fig4}), the interplay between three-body forces in the \nopieft exhibits large effect which completely removes overbinding presented for the Phen-2B interaction in Fig.~\ref{fig5}, yielding correct $B_\Lambda ({\rm ^5_\Lambda He})$, exact $B_\Lambda ({\rm ^3_\Lambda H})$, $B_\Lambda ({\rm ^4_\Lambda H})$, and $E_{\rm exc} ({\rm ^4_\Lambda H^*})$ plus unbound $\rm ^3_\Lambda H^*$ as presented in Fig.~\ref{fig3}. This suggests  that the sensitivity of the $\Lambda nn$ system to the three-body $\Lambda NN$ force seems to be relatively small.

\subsection{Stability and error of continuum solutions \label{stability}}
In this subsection, we demonstrate stability and accuracy of our CSM and IACCC resonance solutions for a particular point of the $\Lambda nn$ pole trajectory. More precisely, we use the $\chi$EFT(LO) \nopieft interaction with $\lambda=1.25~{\rm fm^{-1}}$ and the strength of auxiliary three-body interaction $d_\lambda^{I=1,S=1/2}=-24$~MeV. This specific choice was motivated by large $\theta_r = {\rm arctan}(E/2\Gamma)/2$ angle of the corresponding $\rm \Lambda nn$ resonance energy since it can be already challenging to describe such a pole position accurately within the CSM (see the last $\chi$EFT(LO) CSM solution in the right panel (b) of Fig.~\ref{fig4}).

Using the CSM in a finite basis we make sure that our resonant solution is stable and does not change with an increasing number of basis states. Here, we apply the harmonic oscillator (HO) trap technique \cite{SBBM20} with mass scale $m=939$~MeV (\ref{hotrap}) which provides us with an efficient algorithm to select an appropriate, yet not excessively large CSM basis. For a chosen HO trap length $b$ (\ref{hotrap}), this procedure yields stochastically optimized basis of correlated Gaussians with a maximal typical radius which gets larger as the trap becomes more broad. We choose a grid of increasing trap lengths $b_i$ ranging from 20~fm to 80~fm with 2~fm step and using the HO trap technique for each $b_i$, we prepare 31 different basis sets. In the next step, we build the CSM basis for our resonance calculation in the following way: First, we fix correlated Gaussian states obtained for the lowest $b_0=20$~fm trap length. Second, we take the basis states for $b_1=22$~fm leaving out the states which are nearly linear dependent to any of already fixed $b_0$ correlated Gaussians and we merge $b_0$ and $b_1$ basis sets. Next, in the same way, we add correlated Gaussians from the $b_2=24$~fm basis set to already fixed $b_0$ and $b_1$ states. We continue this procedure for all $b_i$ up to certain $b_{\rm max}$ and construct our final CSM basis set. 

\begin{figure}
\includegraphics[width=0.5\textwidth]{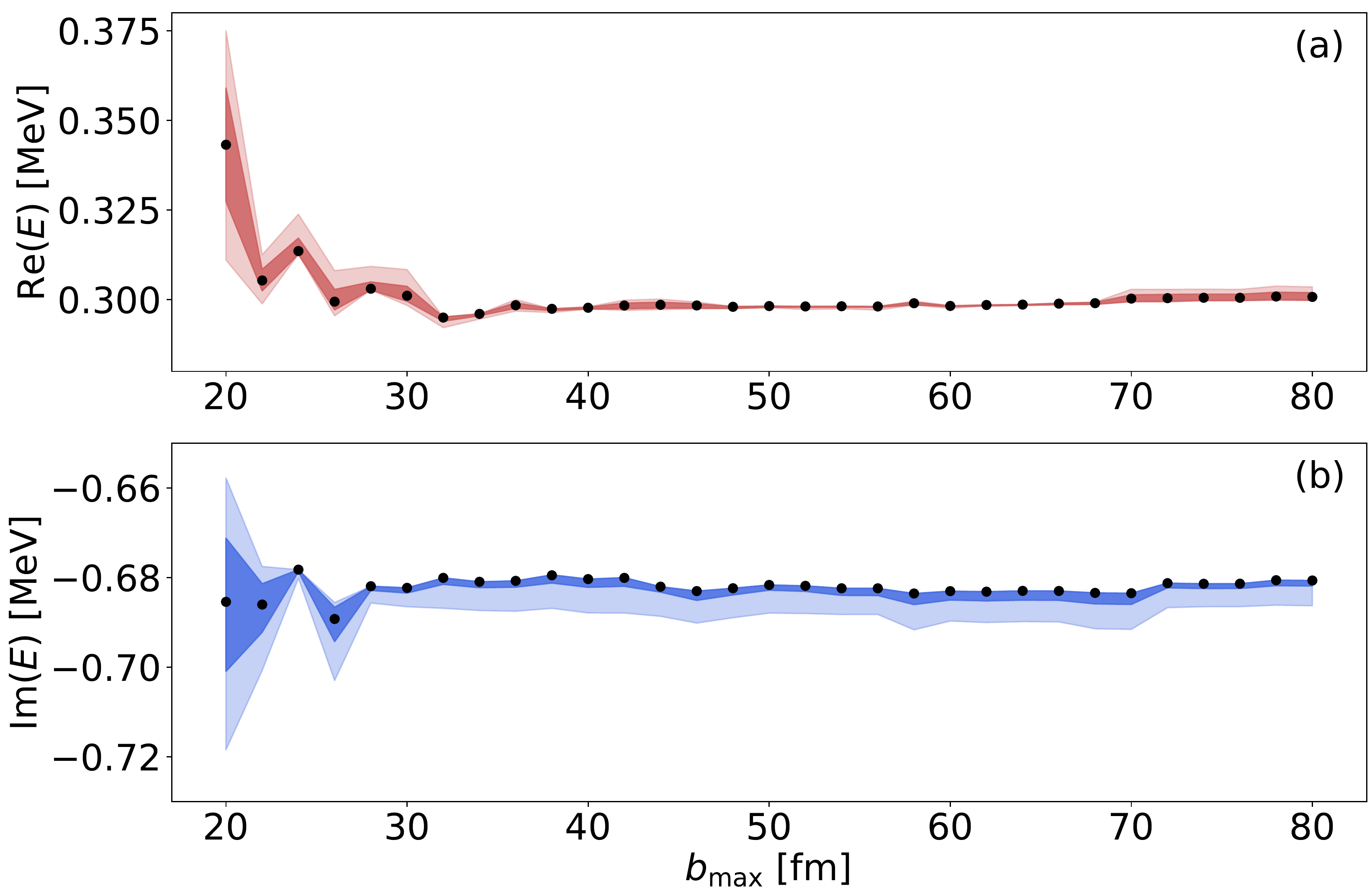}
\caption{\label{fig6} Stability of the $\Lambda nn$ CSM resonant solution $E(\theta)={\rm Re}(E(\theta))+{\rm i}{\rm Im}(E(\theta))$ [upper panel (a): ${\rm Re}(E(\theta))$; lower panel (b): ${\rm Im}(E(\theta))$] as a function of increasing HO trap length $b_{\rm max}$. Black dots show the most stationary point of the $\theta$-trajectory $E(\theta_{\rm opt})$. Darker shaded area shows uncertainty of $E(\theta)$ within $\theta_{\rm opt} \pm 1^\circ$ range, lighter shaded area shows the same within $\theta_{\rm opt} \pm 4^\circ$ range. The particular pole position was calculated for \nopieft interaction with $\chi$EFT(LO) $\Lambda N$ scattering lengths and $\lambda=1.25~{\rm fm^{-1}}$, strength of auxiliary three-body force was set to $d_\lambda^{I=1,S=1/2}=-24~{\rm MeV}$.}
\end{figure}

The stability of the CSM solution with respect to HO trap length $b$ is illustrated in Fig.~\ref{fig6}. Here, we present calculated real and imaginary parts of the $\Lambda nn$ resonance energy using different CSM bases obtained combining HO trap sets up to a certain $b_{\rm max}$. Black dots stand for the most stationary point of the resonance $\theta$-trajectory $E_{\Lambda nn}^{\rm CSM}(\theta_{\rm opt})$ for which $\left|\frac{{\rm d}E}{{\rm d}\theta}\right|_{\theta_{\rm opt}}$ is minimal. Shaded areas then show the spread of resonance energy $E_{\Lambda nn}^{\rm CSM}(\theta)$ within the $\theta_{\rm opt}\pm 1^\circ$ range (darker shaded area) and the $\theta_{\rm opt} \pm 4^\circ$ range (lighter shaded area) thus indicating the level of the CSM resonance energy dependence on the scaling angle $\theta$ (\ref{csmtrans}). The calculated $\Lambda nn$ resonance energy stabilizes already using the CSM basis constructed for $b_{\rm max}=36$~fm. It is clearly visible that considering higher $b_{\rm \max}$ and thus including more basis states does not affect the CSM solution.  

\begin{figure*}
\begin{tabular}{cc}
\includegraphics[width=0.45\textwidth]{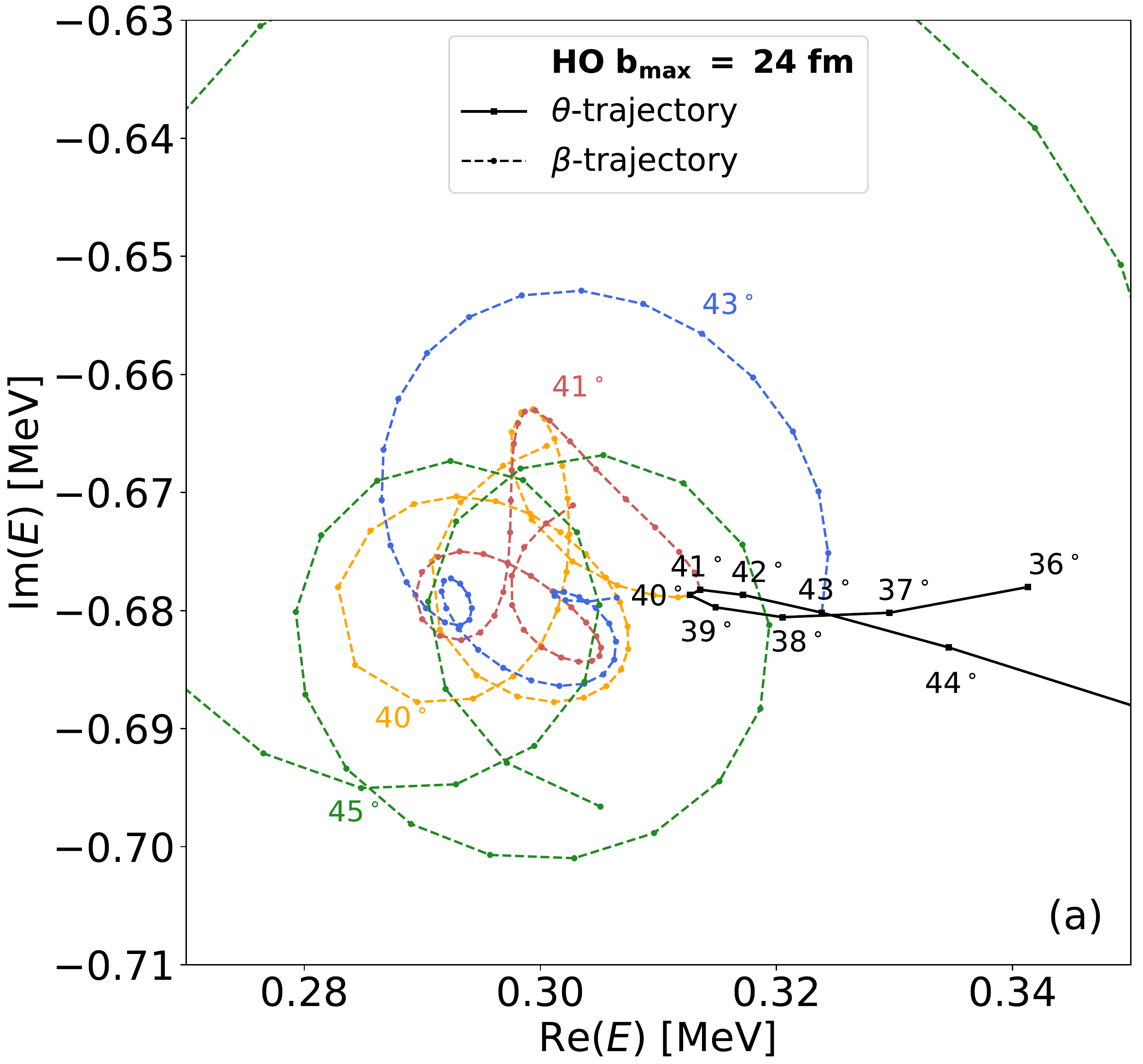}&\includegraphics[width=0.45\textwidth]{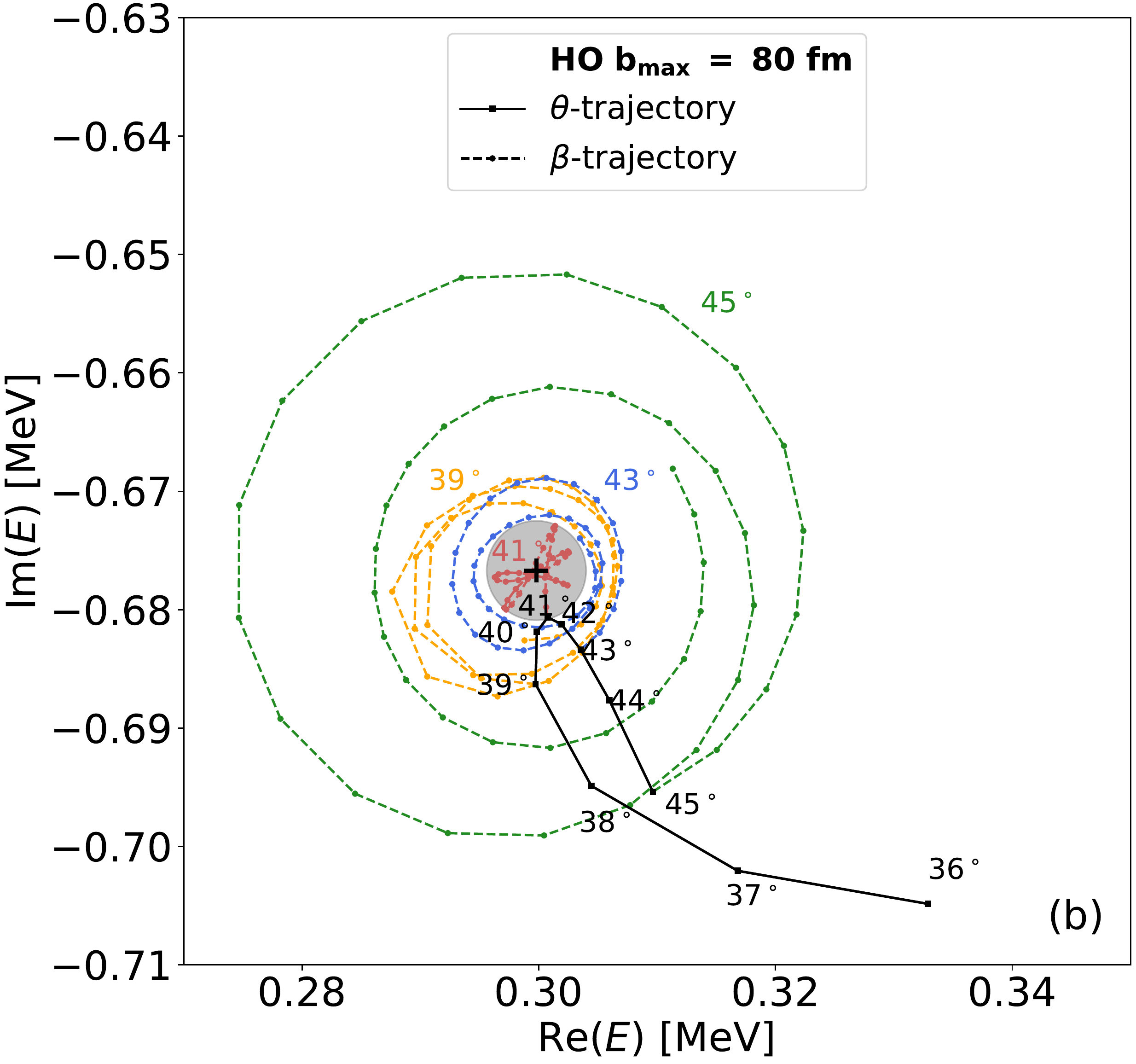}\\
\end{tabular}
\caption{\label{fig7}
$\Lambda nn$ resonance $\theta$-trajectory (Im$(\theta)$=0; black solid line) and $\beta$-trajectories (colored dotted lines) showing movement of corresponding $E(\theta)$ as a function of $\theta$ in the complex energy plane. Trajectories are calculated for two different CSM basis sets which were obtained combining HO trap sets up to $b_{\rm max}=24$~fm (left panel (a)) and up to $b_{\rm max}=80$~fm (right panel (b)). $\beta$-trajectories are presented for several different Re$(\theta)$ changing Im($\theta$) from 0 to 0.44 radians with 0.01 step. Black cross in the left panel indicates estimated $\Lambda nn$ resonance position of the true CSM solution satisfying Eq.~(\ref{statcond}). Shaded gray area then shows corresponding CSM error. $\Lambda nn$ calculation is performed using the same interaction as in Fig.~\ref{fig6}}.
\end{figure*}

In Fig.~\ref{fig7} we show the calculated $\theta$-trajectory and several $\beta$-trajectories for two different CSM bases which were obtained for $b_{\rm max}=24$~fm (left panel (a)) and for $b_{\rm max}=80$~fm (right panel (b)). For $b_{\rm max}=24$~fm we can clearly see that $\beta$-trajectories are not circular and manifest highly unstable behaviour due to poor quality of the employed basis set. In fact, we have already pointed out in Fig.~\ref{fig6} that the $\Lambda nn$ resonance solution stabilizes at least for $b_{\rm max}=36$~fm. Using the CSM basis for $b_{\rm max}=80$~fm (right panel) our results are stable showing almost circular $\beta$-trajectories characterised by their decreasing radius as the corresponding Re$(\theta)$ approaches Re$(\theta_{\rm opt})\approx 41^\circ$. The $\beta$-trajectory for Re$(\theta)=41^\circ$ exhibits oscillatory behavior within a small region around the true CSM solution. We assume that this effect is related to a finite dimension of our CSM basis set and corresponding circular trajectory would be recovered by considering more basis states. The most probable $\Lambda nn$ resonance energy $E_{\Lambda nn}^{\rm CSM}$ is in the center of the grey shaded circle while its radius defines the error of our true CSM solution. In this particular case, the $\Lambda nn$ resonance energy is $E_{\Lambda nn}^{\rm CSM}=0.2998(42)~-~{\rm i}~0.6767(42)~{\rm MeV}$.

The stability of the IACCC solution is demonstrated in Table~\ref{tab2} where we present $\Lambda nn$ resonance energies $E_{\Lambda nn}^{\rm IACCC}$ using different degrees $(M,N)$ of the Pad\'{e} approximant $\mathcal{P}^{(M,N)}$ (\ref{padeapprox}). As expected, calculated $E_{\rm \Lambda nn}^{\rm IACCC}$ start to stabilize with increasing $(M,N)$. The IACCC solution saturates already for (7,7) and does not improve dramatically with further increase of $(M,N)$. This is predominantly explained by finite precision of our SVM bound state energies which are used to fix the parameters of $\mathcal{P}^{(M,N)}$ and by numerical instabilities which slowly start to affect our IACCC solution at higher degrees of the approximant. Comparing saturated IACCC solution obtained with different $(M,N)$ ranging from (7,7) up to (13,13) we estimate for this specific example the $E_{\Lambda nn}^{\rm IACCC}$ accuracy $\sim 3 \times 10^{-3}$~MeV. Despite considerable difference between IACCC and CSM, both approaches predict remarkably consistent $ \Lambda nn$ resonance energies. In fact, all presented IACCC energies starting from the Pad\'{e} approximant of degree (7,7) and higher lie within the errors of the corresponding CSM prediction.
 \begin{table*}
    \centering
    \caption{\label{tab2} Stability of the $\Lambda nn$ resonance energy $E^{\rm IACCC}_{\Lambda nn}$ and $\rm ^3_\Lambda H^*$ virtual state energy with respect to the $\Lambda+d$ threshold $E_{{\rm v},~{\rm ^3_\Lambda H^*}}^{\rm IACCC}$ calculated within the IACCC for increasing degree $(M,N)$ of the Pad\'{e} approximant. $\Lambda nn$ calculation is performed using the same interaction as in Fig.~\ref{fig6}. Position of the $\rm ^3_\Lambda H^*$ virtual state is determined for \nopieft interaction with $\chi$EFT(LO) $\Lambda N$ scattering lengths and $\lambda=1.25~{\rm fm^{-1}}$ with no auxiliary three-body force, i.e. $d_\lambda^{I=0,S=3/2}=0~{\rm MeV}$. $E_{\rm diff}$ stands for the difference between absolute values of IACCC solution calculated for two neighbouring Pad\'{e} approximants $E_{\rm diff}^{(M,N)}=|E^{(M,N)}|-|E^{(M-1,N-1)}|$. All energies are given in MeV.}
    \begin{ruledtabular}
    \begin{tabular}{cccccccc}
        $(M,N)$& $E^{\rm IACCC}_{\Lambda nn}$&$|E^{\rm IACCC}_{\Lambda nn}|$&$E_{\rm diff}$($ \Lambda nn$)&$E_{{\rm v},~{\rm ^3_\Lambda H^*}}^{\rm IACCC}$&$E_{\rm diff}$($\rm ^3_\Lambda H^*$)\\\hline
        (3,3)&-0.0588~-~{\rm i}0.5605&0.5636&&-0.04216&\\
        (4,4)&0.3367~-~{\rm i}0.7041&0.7805&0.2169&-0.05192&0.00976\\
        (5,5)&0.2965~-~{\rm i}0.6559&0.7198&-0.0652&-0.05154&-0.00038\\
        (6,6)&0.2941~-~{\rm i}0.6770&0.7381&0.0183&-0.05161&0.00007\\
        (7,7)&0.3003~-~{\rm i}0.6796&0.7430&0.0050&-0.05160&-0.00001\\
        (8,8)&0.2997~-~{\rm i}0.6796&0.7427&-0.0003&-0.05160&$<10^{-5}$\\
        (9,9)&0.3001~-~{\rm i}0.6796&0.7429&0.0002&-0.05156&-0.00004\\
        (10,10)&0.3014~-~{\rm i}0.6791&0.7430&0.0001&-0.05159&0.00003\\
        (11,11)&0.3012~-~{\rm i}0.6795&0.7433&0.0003& 0.05160&0.00001\\
        (12,12)&0.3020~-~{\rm i}0.6757&0.7401&-0.0032& -0.05160 & $<10^{-5}$\\
        (13,13)&0.3026~-~{\rm i}0.6765&0.7411&0.0010&-0.05161&0.00001\\
        \end{tabular}
        \end{ruledtabular}
\end{table*}

Dependence of our IACCC calculations of the $\rm ^3_\Lambda H^*$ virtual state energy $E_{{\rm v},~{\rm ^3_\Lambda H^*}}^{\rm IACCC}$ on different degrees of the Pad\'{e} approximant is demonstrated in Table~\ref{tab2} as well. In this particular case we use as an example the \nopieft interaction with the $\chi$EFT(LO) $\Lambda N$ scattering lengths, cut-off $\lambda=1.25~{\rm fm^{-1}}$, and no auxiliary interaction. We see that the $\rm ^3_\Lambda H^*$ solution starts to stabilize already for $\mathcal{P}^{(4,4)}$ and it is approximately by two orders more accurate than the solutions for the $\Lambda nn$ resonance. The reason is that the $\rm ^3_\Lambda H^*$ virtual state lies in the vicinity of the $\Lambda+d$ threshold, analytical continuation from the bound region is thus not performed far into the continuum, which enhances the IACCC precision.

The uncertainty of our IACCC resonance solutions in the fourth quadrant of a complex energy plane (${\rm Re}(E)>0$, ${\rm Im}(E)<0$) does not exceed $\approx 4 \times 10^{-3}~{\rm MeV}$. All IACCC results are crosschecked by the CSM in a region of its applicability determined by the maximal resonance angle $\theta_r\approx 35^\circ$ for which our complex scaling results are still reliable. Up to this point the CSM solution possesses the same minimal accuracy as the IACCC solution, however, for higher $\theta_r$ approaching the limiting value $45^\circ$ the CSM solution quickly starts to deteriorate due to numerical instabilities. 

Subthreshold resonance positions are calculated within the IACCC method. For poles residing deeper in this region of a complex energy plane (${\rm Re}(E)<0$, ${\rm Im}(E)<0$)
the precision of our results, predominantly of the imaginary part ${\rm Im}(E)$, decreases. For ${\rm Re}(E) \in (-0.25,0)~{\rm MeV}$ the maximal error of ${\rm Im}(E)$ is $\approx 5\times10^{-3}~{\rm MeV}$, for ${\rm Re}(E) \in (-0.5,-0.25)~{\rm MeV}$ it is $\approx 0.03~{\rm MeV}$, and for ${\rm Re}(E) \in (-1.0,-0.5)~{\rm MeV}$ it is $\approx 0.1~{\rm MeV}$. Since we are primarily interested in a possible experimental observation, i.e. resonance solutions close to or in the fourth quadrant, we deem such accuracy satisfactory, not affecting our conclusions.

The IACCC method proved to be highly precise in the study of near-threshold virtual state positions. Here, we reach accuracy up to $\approx 10^{-4}~{\rm MeV}$ in all considered cases.

\section{Conclusions}
In the present work, we have studied few-body hypernuclear systems $\Lambda nn$ and ${\rm ^3_\Lambda H^*} (J^\pi=3/2^+,~I=0)$ within a LO \nopieft with 2- and 3-body regulated contact terms. The $\Lambda N$ LECs were associated with $\Lambda N$ scattering lengths given by various interaction models and the $\Lambda NN$ LECs were fitted to known $\Lambda$ separation energies $B_{\Lambda}$ in $A\leq 4$ hypernuclei and the excitation energy $E_{\rm exc}({\rm ^4_{\Lambda}H^*})$. 
Few-body wave functions were described within a correlated Gaussians basis. Bound state solutions were obtained using the SVM. The continuum region was studied by employing two independent methods - the IACCC and CSM. 
Our LO \nopieft approach, which accounts for known $s$-shell hypernuclear data, represents a unique tool to describe  within a unified interaction model 3-, 4-, 5- and 6-body hypernuclar systems -- single- and double-$\Lambda$ hypernuclei including continuum states. In that it differs from other similar studies which focused solely on few particular hypernuclei. Moreover, the \nopieft approach allows us to develop systematically higher orders corrections, assess reliably precision of calculations and evaluate errors of their solutions. 

The additional auxiliary 3-body potential introduced to study $\Lambda nn$ and  ${\rm ^3_\Lambda H^*}$ continuum states allows us to explore different scenarios. Fixing the attractive strength of the auxiliary force in order to get these systems just bound yields considerable discrepancy between calculated and experimental  $B_\Lambda$s of 4- and 5-body $s$-shell hypernuclei. Our conclusions thus ruled out the possibility for the existence of bound $\Lambda nn$ and $\rm ^3_\Lambda H^*$ states, which is in accord with conclusions of previous theoretical studies~\cite{DD59,garcilazo87,MKGS95,GFV07a,GFV07b,HOGR14,GG14}. Moreover, we found that by increasing the strength of the $\Lambda$ attraction, the onset of the 
$\rm ^3_\Lambda H^*$ comes before the $\Lambda nn$ binding. 
The experimental evidence for the bound $\Lambda nn$ state reported by the HypHI collaboration~\cite{HypHI13} would thus imply existence of the bound state $\rm ^3_\Lambda H^*$.  

On the basis of our \nopieft calculations with the auxiliary force set to zero, we firmly conclude that the excited state ${\rm ^3_\Lambda H^*}$ is a virtual state. On the other hand, the $\Lambda nn$ pole located close to the three-body threshold in a complex energy plane could convert to a true resonance with Re$(E)>0$ for some considered $\Lambda N$ interactions [e.g., for NSC97f and $\chi$EFT(NLO)] but most likely does not exceed $E_r \approx 0.3~{\rm MeV}$. However, its width $\Gamma$ is rather large -- $1.16 \leq \Gamma \leq 2.00~{\rm MeV}$. Even larger width would be obtained for a rather weak $\Lambda N$ interaction strength but it does not yield experimentally observable $\rm \Lambda nn$ pole. On the contrary, the observation of a sharp resonance would definitely attract considerable attention since it would signal that the $\Lambda N$ interaction at low-momenta is stronger than most $\Lambda N$ interaction models suggest.  

Besides the model dependence of our calculations we explored the stability of solutions with respect to the cutoff parameter $\lambda$. We demonstrated that already for $\lambda =4$~fm$^{-1}$ the calculated energies stabilize close to the asymptotic value corresponding to the renormalization scale invariance limit $\lambda\rightarrow \infty$. We anticipate that the truncation error, describing effects of higher order corrections, is about 47\% and does not change our results qualitatively. In a region accessible by the CSM we performed comparison of the CSM with IACCC method, which yielded highly consistent solutions, hence proving reliability of our results. Moreover, we verified that our CSM solutions for $\Lambda nn$ are stable with respect to the considered number of basis states. Exploring both the $\theta$ and $\beta$ trajectories of the $\Lambda nn$ pole for one particular case we set the true CSM solution including its error. The stability of the IACCC method  with respect to the degree of the employed  Pad\'{e} approximant was investigated and the uncertainty of the calculations was assessed. 

A rather different situation occurs when we consider just 2-body phenomenological interactions fitted to $NN$ and $\Lambda N$ scattering lengths and effective ranges. We then obtain subthreshold $\Lambda nn$ pole positions close to those of Afnan and Gibson \cite{AG15}. However, these interactions fail to describe other few-body $\Lambda$ hypernuclei. The predicted overbinding of the $s$-shell hypernuclei induced by these phenomenological 2-body interactions indicates a missing repulsive part of the $\Lambda$ interaction. In the \nopieft, it is provided by an additional $\Lambda NN$ 3-body force. A comparison with our LO \nopieft calculations revealed that the results of Afnan and Gibson could be reproduced for the finite cutoff value $\lambda_s \approx 1.25~{\rm fm^{-1}}$. However, thanks to the repulsive $\Lambda NN$ force the $s$-shell hypernuclear data are now described successfully. The LO \nopieft with  $\lambda_s \approx 1.25~{\rm fm^{-1}}$ could thus be considered as a suitable phenomenological model. 

Our method presented here can be directly applied to the double-$\Lambda$ hypernuclear continuum using the recently introduced $\Lambda \Lambda$ extension of a LO \nopieft \cite{CSBGM19}. It is highly desirable to explore possible resonances in the neutral $\Lambda\Lambda n$ and $\Lambda\Lambda nn$ systems or in the $\rm ^4_{\Lambda \Lambda} H$ hypernucleus, where a consistent theoretical continuum study has not been performed yet. Indeed, an example of its importance is the continuing  ambiguity in interpretation of the AGS-E906 experiment \cite{ahn01} referred to as the E906 puzzle. It was firstly interpreted as the bound $\rm ^4_{\Lambda \Lambda} H$ system \cite{ahn01}, however, more recent analyses suggested that the decay of the $\rm ^{~7}_{\Lambda \Lambda} He$ \cite{RH07} or $\Lambda \Lambda nn$ \cite{BBGP19} hypernucleus might provide more plausible interpretation.

This clearly demonstrates the growing importance of precise few-body continuum studies which, although being difficult to conduct, significantly contribute to the complete picture of a stability of  hypernuclear systems. In fact, the applicability of our few-body approach is rather broad in principle -- it might be used not only to calculations of  hypernuclear systems but also  $\eta$ or $K^-$ mesic nuclei, or even atoms.

\begin{acknowledgments}
We are grateful to Avraham Gal for valuable discussions and careful reading of the manuscript. 
This work was partly supported by the Czech Science Foundation GACR grant 19-19640S.
The work of NB was supported by the Pazy Foundation and by the Israel Science Foundation grant 1308/16. Furthermore, MS and NB were partially funded by the European Union’s Horizon 2020 research and innovation programme under grant agreement No. 824093. 
\end{acknowledgments}

\end{document}